\documentclass[aps,showpacs,preprintnumbers,amsmath,amssymb,nofootinbib, twocolumn]{revtex4}
\usepackage{graphicx}
\usepackage{mathrsfs}

\begin{document}
\title{Can accretion disk properties observationally distinguish black holes from naked singularities?}
\author{Z. Kov\'{a}cs}
\email{zkovacs@hku.hk}
\address{Department of Physics and Center for Theoretical and Computational Physics, The University of Hong Kong, Pok Fu Lam Road, Hong Kong, P. R. China}
\author{T. Harko}
\email{harko@hkucc.hku.hk}
\address{Department of Physics and Center for Theoretical and Computational Physics, The University of Hong Kong, Pok Fu Lam Road, Hong Kong, P. R. China}

\begin{abstract}
Naked singularities are hypothetical astrophysical objects, characterized by a gravitational singularity without an event horizon. Penrose has proposed a conjecture, according to which there exists a cosmic censor who forbids the occurrence of naked singularities. Distinguishing between astrophysical black holes and naked singularities is
a major challenge for present day observational astronomy. In the context of stationary and axially symmetrical geometries, a possibility of differentiating naked singularities from black holes is through the comparative study of thin accretion disks properties around rotating naked singularities and Kerr-type black holes, respectively. In the present paper, we consider accretion disks around axially-symmetric rotating naked singularities, obtained as solutions of the field equations in the Einstein-massless scalar field theory. A first major difference between rotating naked singularities and Kerr black holes is in the frame dragging effect, the angular velocity of a rotating naked singularity being inversely proportional to its spin parameter. Due to the differences in the exterior geometry, the thermodynamic and electromagnetic properties of the disks (energy flux, temperature distribution and equilibrium radiation spectrum) are different for these two classes of compact objects, consequently giving clear observational signatures that could discriminate between black holes and naked singularities. For specific values of the spin parameter and of the scalar charge, the energy flux from the disk around a rotating naked singularity can exceed by several orders of magnitude the flux from the disk of a Kerr black hole.  In addition to this, it is also shown that the conversion efficiency of the accreting mass into radiation by rotating naked singularities is always higher than the conversion efficiency for black holes, i.e., naked singularities provide a much more efficient mechanism for converting mass into radiation than black holes. Thus, these observational signatures may provide the necessary tools from clearly distinguishing rotating naked singularities from Kerr-type black holes.

\end{abstract}
\date{\today}

\maketitle

\section{Introduction}

Investigating the final fate of the gravitational collapse of an initially
regular distribution of matter, in the framework of the Einstein theory of
gravitation, is one of the most active fields of research in contemporary
general relativity. One would like to know whether, and under what initial
conditions, gravitational collapse results in black hole formation. One
would also like to know if there are physical collapse solutions that lead
to naked singularities. If found, such solutions would be counterexamples of
the cosmic censorship hypothesis, which states that curvature singularities
in asymptotically flat space-times are always shrouded by event horizons.

Roger Penrose \cite{Pe69} was the first to propose the idea, known as cosmic
conjecture: does there exist a cosmic censor who forbids the occurrence of
naked singularities, clothing each one in an absolute event horizon? This
conjecture can be formulated in a strong sense (in a reasonable space-time
we cannot have a naked singularity) or in a weak sense (even if such
singularities occur they are safely hidden behind an event horizon, and
therefore cannot communicate with far-away observers). Since Penrose' s
proposal, there have been various attempts to prove the conjecture (see \cite
{Jo93} and references therein). Unfortunately, no such attempts have been
successful so far.

Since, due to the complexity of the full Einstein equations, the general
problem appears intractable, metrics with special symmetries are used to
construct gravitational collapse models. One such case is the
two-dimensional reduction of general relativity obtained by imposing
spherical symmetry. Even with this reduction, however, very few
inhomogeneous exact nonstatic solutions have been found \cite{Va51}-\cite{Or98}.

Within the framework of various physical models, the spherical gravitational
collapse has been analyzed in many papers \cite{SiJo96}-\cite{No99}. The
idea of probing naked space-times singularities with waves rather than with
particles has been proposed in \cite{IsHo99}. For some space-times the
classical singularity becomes regular if probed with waves, while stronger
classical singularities remain singular.

In order to obtain the energy-momentum tensor for the collapse of a null
fluid an inverted approach was proposed in \cite{Hu96}.  In the framework of the same approach a large class of solutions, including Type II fluids, and which includes most of the known solutions of
the Einstein field equations, has been derived
\cite{WaWu98}, \cite{Ro02}.
The Vaidya radiating metric has been extended to include both a radiation
field and a string fluid in \cite{GlKr98, GlKr99, GoGo03}. The collapse of the quark fluid, described
by the bag model equation of state $p=\left( \rho -4B\right) /3$, with $B=$%
constant, has been studied in \cite{HaCh00}, and the
conditions for the formation of a naked singularity have been obtained. The
obtained solution has been generalized to arbitrary space-time dimensions
and to a more general linear equation of state in \cite
{GhDh02,GhDh03}. The gravitational collapse of a high-density null charged matter fluid, satisfying the Hagedorn equation of state, was considered in the framework of the Vaidya geometry in \cite{ha}.

Since theoretical studies alone cannot give an answers about the existence or non-existence of naked singularities in nature, the differences in the observational properties of the black holes and naked singularities could be used to discriminate between these two classes of objects. Such a distinctive observational feature is represented by the lensing properties, and  a research project that uses gravitational lensing as a tool to differentiate between naked singularities and black holes was initiated in \cite{vir1}. This pioneering work was extended in \cite{vir2}, where the distinctive lensing features of black holes and naked singularities were studied in detail. The gravitational lensing due to a strongly naked singularity is qualitatively different from that due to a Schwarzschild black hole; a strongly naked singularity gives rise to either two or nil Einstein ring(s), and one radial critical curve. The gravitational lensing (particularly time delay, magnification centroid, and total magnification) for a Schwarzschild black hole and for a Janis-Newman-Winicour naked singularities \cite{jnw} were studied in \cite{vir3}. The lensing features are qualitatively similar (though quantitatively different) for Schwarzschild black holes, weakly naked, and marginally strongly naked singularities. However, the lensing characteristics of strongly naked singularities are qualitatively very different from those due to Schwarzschild black holes. Gravitational lensing by rotating naked singularities was considered in \cite{bul},   and it was shown that the shift of the critical curves as a function of the lens angular momentum decreases slightly for the weakly naked, and vastly for the strongly naked singularities, with the increase of the scalar charge.

It is the purpose of the present paper to consider another observational possibility that may distinguish naked singularities from black holes, namely, the study of the properties of the thin
accretion disks around rotating naked singularities and black holes,
respectively. Thus, in the following we consider a comparative study of the physical properties of thin accretion disks around a particular type of rotating naked singularity, and rotating black holes, described by the Kerr metric, respectively. In particular, we consider the basic physical parameters describing the disks, like the emitted energy flux, the temperature distribution on the surface of the disk, as well as the spectrum of the emitted equilibrium radiation. Due to the differences in the exterior geometry, the thermodynamic and electromagnetic properties of the disks (energy flux, temperature distribution and equilibrium radiation spectrum) are different for these two classes of compact objects, thus giving clear observational signatures, which may allow to discriminate, at least in principle, naked singularities from black holes. We would like to point out that the proposed method for the detection of the naked singularities by studying accretion disks is an {\it indirect} method, which must be complemented by {\it direct} methods of observation of the "surface" of the considered black hole/naked singularity candidates, and/or by  the study of the lensing properties of the central compact object.

The mass
accretion around rotating black holes was studied in general
relativity for the first time in \cite{NoTh73}. By using an
equatorial approximation to the stationary and axisymmetric
spacetime of rotating black holes, steady-state thin disk models
were constructed, extending the theory of non-relativistic
accretion \cite{ShSu73}. In these models hydrodynamical
equilibrium is maintained by efficient cooling mechanisms via
radiation transport, and the accreting matter has a Keplerian
rotation. The radiation emitted by the disk surface was also
studied under the assumption that black body radiation would
emerge from the disk in thermodynamical equilibrium. The radiation
properties of thin accretion disks were further analyzed  in
\cite{PaTh74,Th74}, where the effects of photon capture by the
hole on the spin evolution were presented as well. In these works
the efficiency with which black holes convert rest mass into
outgoing radiation in the accretion process was also computed.
More recently, the emissivity properties of the accretion disks
were investigated for exotic central objects, such as wormholes, non-rotating or rotating quark, boson or
fermion stars, brane-world black holes, $f(R)$ type gravity models, and Horava-Lifshitz gravity
\cite{Harko}. In all these cases
it was shown that particular signatures can appear in the
electromagnetic spectrum, thus leading to the possibility of
directly testing different physical models by using astrophysical
observations of the emission spectra from accretion disks.

The present paper is organized as follows. The geometry of the considered naked singularity is presented in Section \ref{0}. In Section \ref{2} we obtain the main physical parameters (specific energy, the specific angular momentum, and angular velocity) for massive test particles in stable circular orbits in stationary and axisymmetric spacetimes. The motion of test particles around rotating naked singularities is considered in Section \ref{mot}. The frame dragging effect is analyzed in Section \ref{mot1}. The properties of standard thin accretion disks are reviewed in Section \ref{std}. The energy flux, temperature distribution, and radiation spectrum from thin disks around naked singularities and Kerr black holes are discussed in Section {\ref{el}. Some observational implications of our results are considered in Section \ref{impl}. We discuss and conclude our results in Section \ref{dis}.

\section{Spacetime geometry of the naked singularity}\label{0}

As an example of a rotating naked singularity geometry we will consider in the following the Kerr-like solution of the Einstein gravitational field equations with a massless scalar field, $R_{\mu \nu}=-8\pi \varphi _{,\mu}\varphi _{,\nu }$, where $\varphi $ is the scalar field, obtained in \cite{sol}.  In the coordinate system $(t,r,\theta,\phi)$, the line element, adapted to the axial symmetry, is given by
\begin{eqnarray}
ds^2&=&-f^{\gamma}(dt-wd\phi)^2-2w(dt-wd\phi)d\phi\nonumber\\
& & + f^{1-\gamma} \Sigma (\Delta^{-1}dr^2+d\theta^2+\sin^2\theta d\phi^2)\;,\label{ds2}
\end{eqnarray}
where
\begin{eqnarray*}
f(r,\theta)&=&1-2\frac{Mr}{\gamma\Sigma}\;,\\
w(\theta)&=&a\sin^2\theta\:,\\
\Sigma(r,\theta)&=&r^2+a^2\cos^2\theta\:,\\
\Delta(r)&=&r^2+a^2-2\frac{Mr}{\gamma}\;,
\end{eqnarray*}
 and $\gamma $ is a constant.

 The energy-momentum tensor of the massless scalar field generating the naked singularity is given by
 \begin{equation}\label{enmom}
 T_{\mu \nu }^{\varphi }=\varphi _{,\mu }\varphi _{,\nu }-\frac{1}{2}g_{\mu \nu }g^{\alpha
\beta }\varphi _{,\alpha }\varphi _{,\beta }.
\end{equation}
 The scalar field satisfies the equation $g^{\mu \nu}\nabla _{\mu }\nabla _{\nu}\varphi =0$, which follows from the conservation of the energy-momentum tensor, $\nabla ^{\mu}T_{\mu \nu }^{\varphi }=0$,  and is given by
 \begin{equation}\label{scal}
\varphi =\frac{\sqrt{1-\gamma ^{2}}}{4}\ln \left[ 1-\frac{2Mr}{\gamma \left(
r^{2}+a^{2}\cos ^{2}\theta \right) }\right] .
\end{equation}

For $0<\gamma<1$ this metric describes the spacetime geometry of a naked singularity, with a total mass $M$, and an angular momentum $J=aM=a_*M^2$. Here
$a_*=J/M$ is the dimensionless spin parameter. It can be shown that the scalar curvature of this geometry diverges at a radius where the condition $g_{tt}=-f^{\gamma}=0$ holds, which indicates the existence of a curvature singularity at the radius
\begin{equation}
r_s=\gamma^{-1}M(1+\sqrt{1-\gamma^2a^2_*\cos^2\theta})\;.\label{rs}
\end{equation}
Considering the angular dependence of this expression, we see that the minimal
radial coordinate of the singularity is at the pole, $r_s(\theta=0)=\gamma^{-1}M(1+\sqrt{1-\gamma^2a^2_*})$,  whereas it has a maximal radius in the equatorial plane, $r_s(\theta=\pi/2)=2M/\gamma$. The bigger root of the equation $\Delta(r)=0$ is
\begin{equation}
r_{+}=\gamma^{-1}M(1+\sqrt{1-\gamma^{2}a^2_*})\:.\label{rp}
\end{equation}
Since for any value of $\theta$, $r_{+}$ is always less than or equal to $r_s$, the singularity is not hidden behind the event horizon. It can also be shown that there is at least one null geodesic in the equatorial plane connecting the singularity with the future null infinity, i.e., the singularity is naked \cite{proof}.

For $\gamma=1$ the line element describes the spacetime geometry of the Kerr black hole, where the vanishing horizon function, $\Delta(r)=0$, has the roots
$r_{\pm}=M(1\pm \sqrt{1-a^2_*})$. Then $r_+$ gives the horizon radius for $|a_*|<1$. In the case of rotating black holes, the condition $g_{tt}=0$ does not provide any singular surface, but the ergosphere located at $r_e=M(1+\sqrt{1-a^2_*\cos^2\theta})$.

We note that since for the metric (\ref{ds2}) $g_{t\phi}=-w(1-f^{\gamma})$ and $g_{\phi\phi}=f^{1-\gamma}\Sigma\sin^{2}\theta+w^{2}(2-f^{\gamma})$, respectively,  the frame dragging frequency of this rotating solution can be written as
\begin{equation}
\omega =\frac{w(1-f^{\gamma})}{f^{1-\gamma}\Sigma\sin^{2}\theta+w^{2}(2-f^{\gamma})}\;.\label{omega}
\end{equation}
For the Kerr black holes ($\gamma=1$) we obtain the familiar expression $\omega=2MarA^{-1}$, with $A(r,\theta)=(r^2+a^2)-\Delta a^2\sin^2\theta$.

In the equatorial approximation ($|\theta-\pi/2|\ll1$),  the components of the metric (\ref{ds2}) reduce in the equatorial coordinate system $(t,r,z=r\cos\theta,\phi)$ to the form
\begin{eqnarray}
g_{tt} & = & -f^{\gamma}\:,\label{gtt}\\
g_{t\phi} & = & -a(1-f^{\gamma})\:,\\
g_{\phi\phi} & = & f^{1-\gamma}r^{2}+a^{2}(2-f^{\gamma})\:,\label{gpp}\\
g_{rr} & = & f^{1-\gamma}\frac{r^{2}}{\Delta}\:,\\
g_{zz} & = & f^{1-\gamma}\:\label{gzz},\end{eqnarray}
 where
 \[f(r)=1-\frac{2M}{\gamma r}\:.\] The scalar field is given, in this approximation, by $\varphi =$ $\sqrt{1-\gamma ^{2}}\ln f/4$.
In the equatorial plane the frame dragging frequency, given by Eq.~(\ref{omega}), has the expression
\begin{equation}
\omega= 2M^{-1}a_*x^{-6}\mathscr{A}^{-1}\:,\label{omegaeq}
\end{equation}
where $x=\sqrt{r/M}$, and
\[\mathscr{A}=2\frac{f^{1-\gamma}+a^2_*x^{-4}(2-f^{\gamma})}{x^2(1-f^{\gamma})},
\]
respectively, with $f(x)=1-2\gamma^{-1}x^{-2}$. For the Kerr solution ($\gamma =1$), $\mathscr{A}$  reduces to the form \cite{NoTh73,PaTh74}
\[\mathscr{A}=1+a^2_*x^{-4}+2a^2_*x^{-6}.
\]

\section{Circular geodesic motion in stationary and axisymmetric spacetimes}\label{2}

Let us consider, in the coordinate system $(t,r,\theta,\phi)$, an arbitrary stationary and axially symmetric geometry, with line element given by
\begin{equation}
ds^{2}=g_{tt}dt^{2}+2g_{t\phi }dtd\phi +g_{rr}dr^{2}+g_{\theta \theta
}d\theta ^{2}+g_{\phi \phi }d\phi ^{2}\;.  \label{ds2rcoappr}
\end{equation}%
 The metric (\ref{ds2rcoappr}) is adapted to the symmetries of the spacetime, endowed with the time- and space-like Killing vectors $(\partial/ \partial t)^{\mu}$ and $(\partial/ \partial \phi)^{\mu}$ for time translations and spatial rotations, respectively. In the equatorial approximation ($|\theta -\pi /2|\ll 1$), and the metric functions $g_{tt}$, $g_{t\phi }$, $g_{rr}$, $g_{\theta
\theta }$ and $g_{\phi \phi }$ in Eq.~(\ref{ds2rcoappr}) depend only on the radial coordinate $r$. Then the geodesic equations in the equatorial plane take the form
\begin{equation}
 \frac{dt}{d\tau }=\frac{\widetilde{E}g_{\phi \phi }+%
\widetilde{L}g_{t\phi }}{g_{t\phi }^{2}-g_{tt}g_{\phi \phi }},
\; \frac{d\phi }{d\tau }=-\frac{\widetilde{E}g_{t\phi }+%
\widetilde{L}g_{tt}}{g_{t\phi }^{2}-g_{tt}g_{\phi \phi }},
\end{equation}%
and
\begin{equation}\label{geodeqs3}
g_{rr}\left( \frac{dr}{d\tau }\right) ^{2}=V(r),
\end{equation}%
respectively, where $\tau $ is the affine parameter, $\widetilde{E}$ and $\widetilde{L}$ are the specific energy and specific angular momentum of the particles moving along the time-like geodesics, and the potential term $V(r)$ is defined by
\begin{equation}
V(r)\equiv -1+\frac{\widetilde{E}^{2}g_{\phi \phi }+2\widetilde{E}%
\widetilde{L}g_{t\phi }+\widetilde{L}^{2}g_{tt}}{g_{t\phi
}^{2}-g_{tt}g_{\phi \phi }}\;.
\end{equation}

For circular orbits in the equatorial plane, the conditions  $V(r)=0$ and $V_{,r}(r)=0$, respectively, must
hold. These conditions determine the specific energy $\widetilde{E}$, the specific angular momentum $\widetilde{L}$ and
the angular velocity $\Omega$ of particles moving on circular orbits as
\begin{eqnarray}
\widetilde{E} &=&-\frac{g_{tt}+g_{t\phi }\Omega }{\sqrt{-g_{tt}-2g_{t\phi
}\Omega -g_{\phi \phi }\Omega ^{2}}}\;,  \label{tildeE} \\
\widetilde{L} &=&\frac{g_{t\phi }+g_{\phi \phi }\Omega }{\sqrt{%
-g_{tt}-2g_{t\phi }\Omega -g_{\phi \phi }\Omega ^{2}}},  \label{tildeL} \\
\Omega  &=&\frac{d\phi }{dt}=\frac{-g_{t\phi ,r}+\sqrt{(g_{t\phi
,r})^{2}-g_{tt,r}g_{\phi \phi ,r}}}{g_{\phi \phi ,r}}\;.\label{Omega}
\end{eqnarray}
Any stationary observer, moving along a world line $r={\rm constant}$ and $\theta={\rm constant}$ with a uniform angular velocity $\Omega$, has a four-velocity vector $u^{\mu}\propto(\partial/\partial t)^{\mu}+\Omega(\partial/\partial{\phi})^{\mu}$, which lies inside the surface of the future light cone. Therefore, the condition
\begin{equation}
 [(\partial/\partial t)^{\mu}+\Omega(\partial/\partial{\phi})^{\mu}]^2
= g_{tt} + 2 \Omega g_{t\phi} + \Omega^2 g_{\phi\phi} \leq 0,\label{cond}
\end{equation}
appearing in Eqs.~(\ref{tildeE}) and (\ref{tildeL}), provides the constraint $\Omega_{min}<\Omega<\Omega_{max}$
for the angular velocity of the stationary observers, with
\begin{eqnarray}
\Omega_{min}&=&\omega-\sqrt{\omega^2-\frac{g_{tt}}{g_{\phi\phi}}}\:,\label{Omin}\\
\Omega_{max}&=&\omega+\sqrt{\omega^2-\frac{g_{tt}}{g_{\phi\phi}}}\:,\label{Omax}
\end{eqnarray}
where the frame dragging frequency of the spacetime is defined by $\omega=-g_{t\phi}/g_{\phi\phi}$.

Only circular orbits for which the condition (\ref{cond}) holds do exist. The limiting case of this condition, $g_{tt} + 2 \Omega g_{t\phi} + \Omega^2 g_{\phi\phi} = 0$,  gives $r_{ph}$, the innermost boundary of the circular orbits for particles, called photon orbit. The circular orbits $(r>r_{ph})$, for which the condition $\widetilde{E}<1$ holds, are bound, and the condition $\widetilde{E}=1$ gives the radius $r_{mb}$ of the marginally bound circular orbit, i.e., the innermost orbits. The marginally stable circular orbits $r_{ms}$ around the central object are determined by the condition
\begin{widetext}
\begin{equation}
V_{,rr}|_{r=r_{ms}}=\left.\frac{\widetilde{E}^{2}g_{\phi \phi ,rr}+2\widetilde{E}\widetilde{L}g_{t\phi ,rr}+%
\widetilde{L}^{2}g_{tt ,rr}-(g_{t\phi }^{2}-g_{tt}g_{\phi \phi})_{,rr}}{g^2_{t\phi}-g_{tt}g_{t\phi}}\right|_{r=r_{ms}}=0\;,  \label{stable}
\end{equation}
\end{widetext}
where the condition $V_{,rr}<0$ holds for all stable circular orbits. The  marginally stable orbit is the innermost boundary of the stable circular orbits of the Keplerian rotation.
By inserting Eqs.~(\ref{tildeE})-(\ref{tildeL}) into Eq.~(\ref{stable}), and
solving the resulting equation for $r$, we obtain the marginally stable
orbit, once the metric coefficients $g_{tt}$, $g_{t\phi }$ and $%
g_{\phi \phi }$ are explicitly given.

\section{Equatorial geodesic motion around rotating naked singularities}\label{mot}

Inserting the metric components (\ref{gtt})-(\ref{gpp}) into Eqs.~(\ref{tildeE})-(\ref{Omega}), we obtain the specific energy, the specific angular momentum, and the angular velocity of the particles orbiting along circular geodesics in the equatorial plane of the rotating naked singularity. Hence $\Omega$, $\widetilde{E}$ and ${\widetilde{L}}$ can be written as
\begin{eqnarray}
\Omega&=&M^{-1}x^{-3}\mathscr{B}^{-1},\label{Omega2}\\
\widetilde{E}&=&\mathscr{C}^{-1/2}\mathscr{G},\label{tildeE2}\\
\widetilde{L}&=&Mx\mathscr{C}^{-1/2}\mathscr{F},\label{tildeL2}
\end{eqnarray}
where
\begin{eqnarray*}
\mathscr{B}&=&f^{1-\gamma}h+a_*x^{-3}\:,\label{calB}\\
\mathscr{C}&=&f^{1-\gamma}[ (fh+2a_{*}x^{-3})h-x^{-2}] ,\\
\mathscr{G}&=&fh+a_{*}x^{-3},\\
\mathscr{F}&=&f^{1-\gamma}[1-a_{*}\left(1-f^{\gamma}\right)x^{-1}h]+a_{*}^{2}x^{-4},
\end{eqnarray*}
with $h(x)=[1-(1-\gamma)(2-\gamma x^2)^{-1}]^{1/2}$. For $\gamma=1$ we have $f(x)=1-2x^{-2}$ and $h=1$, and the latter formulae reduce to the familiar expressions
\begin{eqnarray*}
\mathscr{B}&=&1+a_* x^{-3},\\
\mathscr{C}&=&1-3x^{-2}+2a_* x^{-3},\\
\mathscr{G}&=&1-2x^{-2}+a_* x^{-3},\\
\mathscr{F}&=&1-2a_* x^{-3}+a^2_* x^{-4},
\end{eqnarray*}
obtained for the equatorial approximation of the Kerr solution \cite{NoTh73,PaTh74}.

If we compare the locations of the marginally stable, of the marginally bound, and of the photon orbits, plotted in Fig.~\ref{fig1} as functions of the spin parameter $a_*$, we see that they have a strong dependence on the parameter $\gamma$.

\begin{widetext}

\begin{figure}
\centering
\includegraphics[width=.48\textwidth]{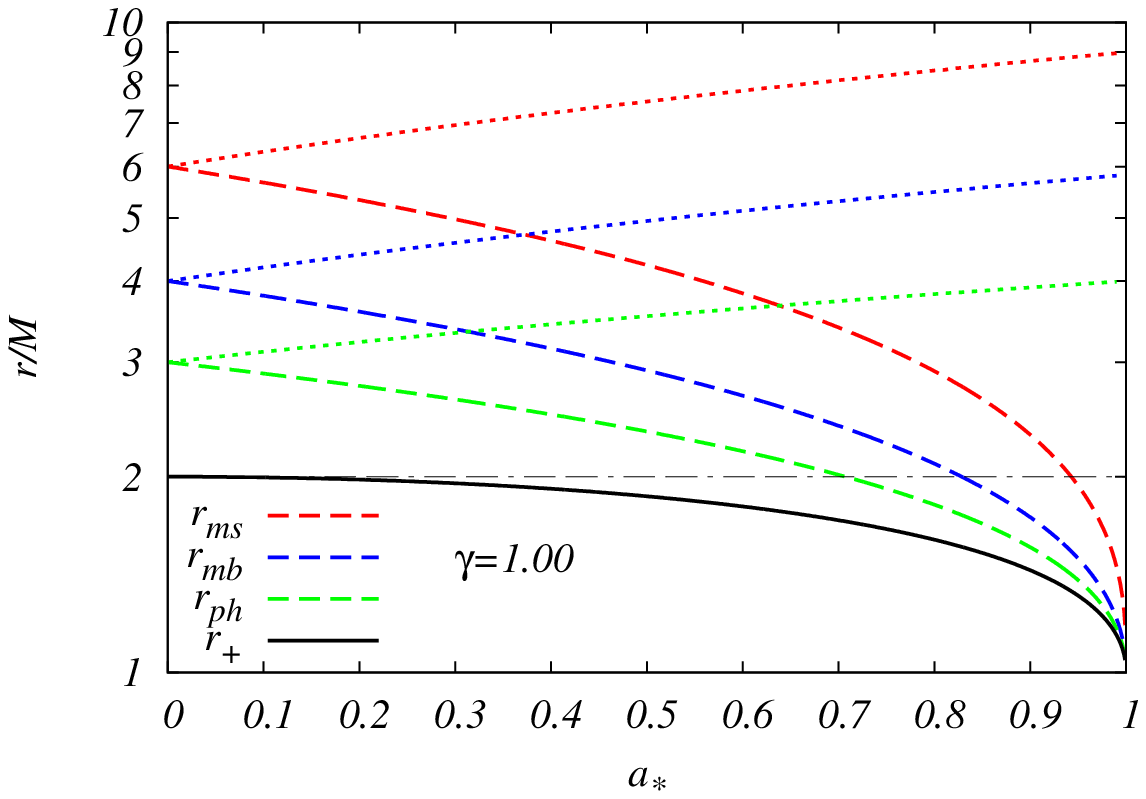}
\includegraphics[width=.48\textwidth]{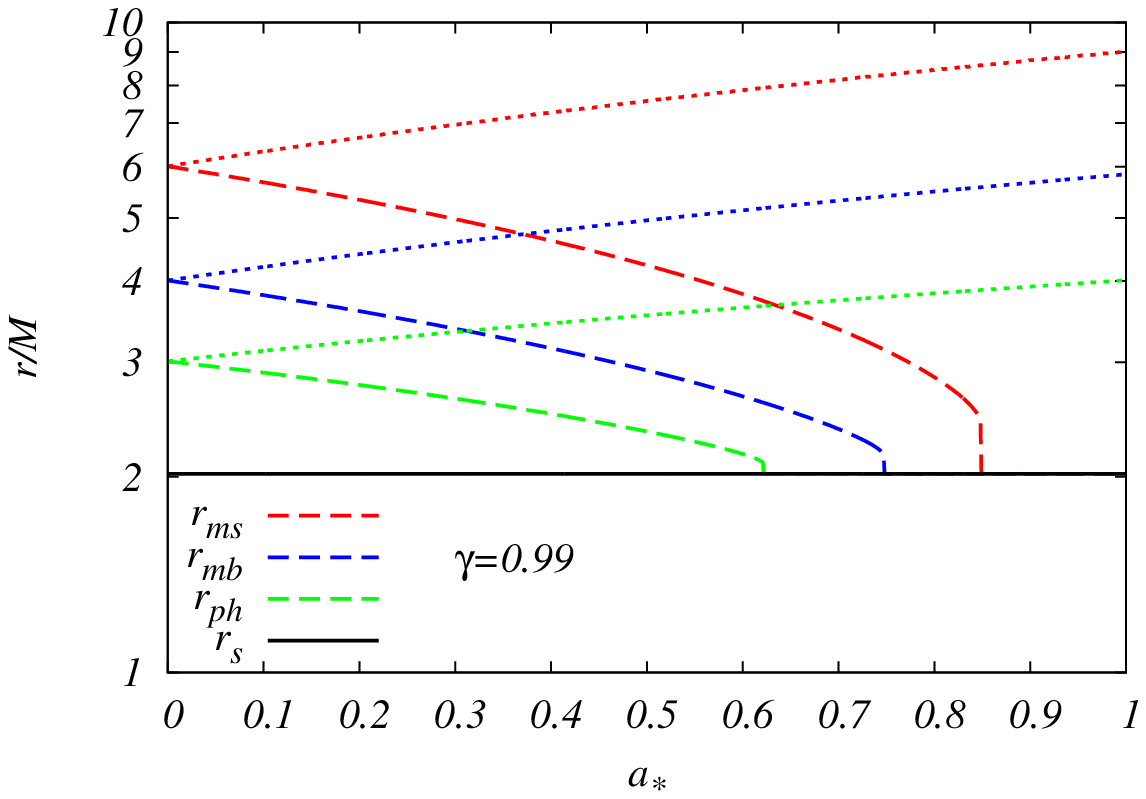}\\
\includegraphics[width=.48\textwidth]{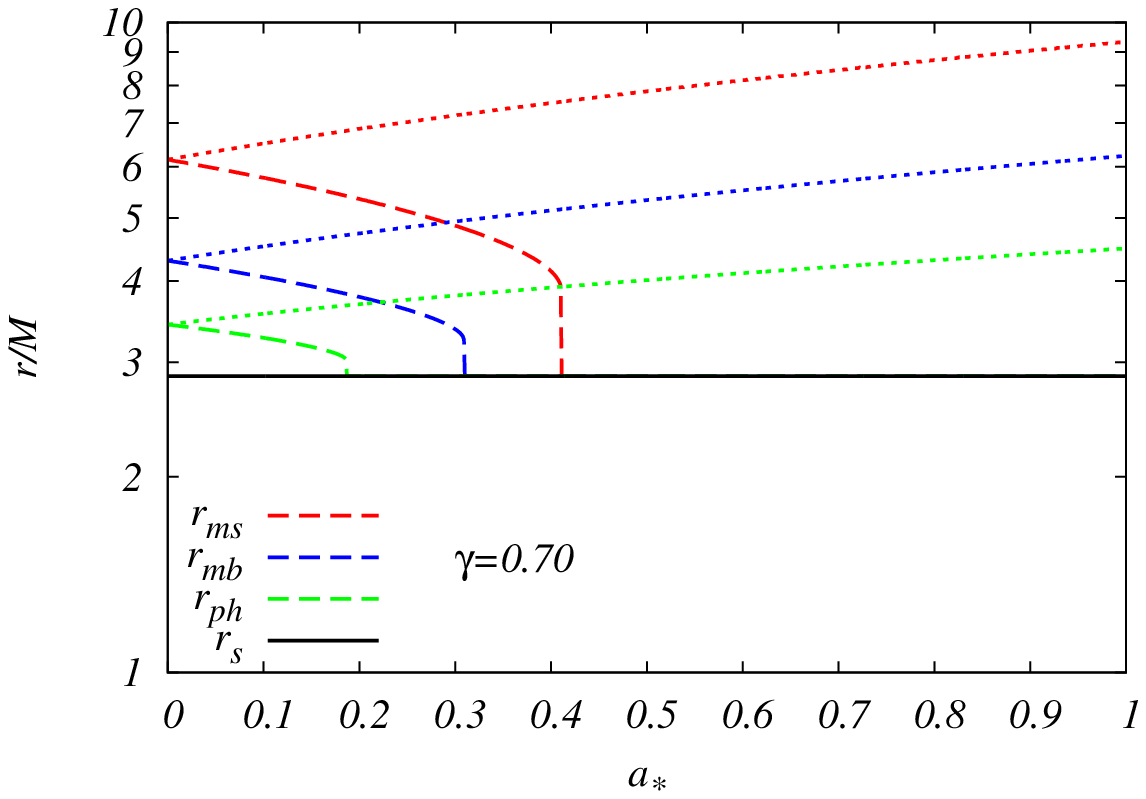}
\includegraphics[width=.48\textwidth]{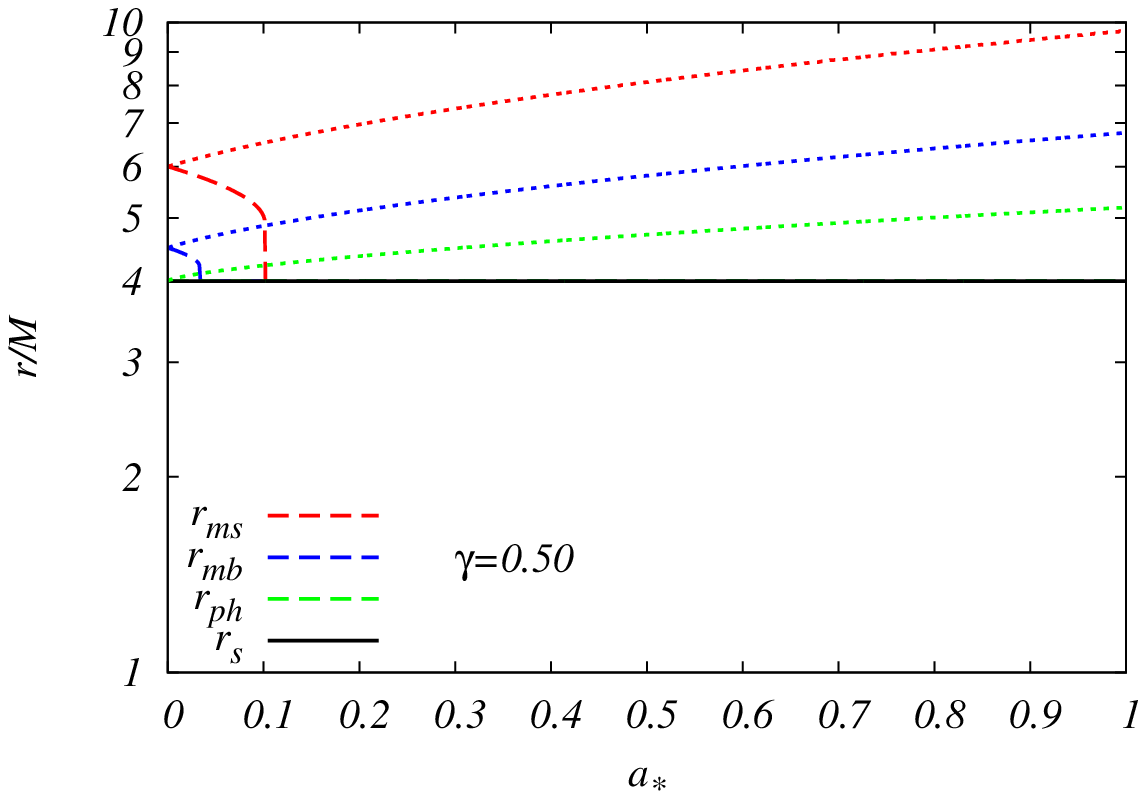}\\
\caption{Radii of circular equatorial orbits around a black hole ($\gamma=1.00$), and a naked singularity ($\gamma=0.99$, 0.7 and 0.5), with a total mass $M$, as functions of the spin parameter $a_*$. The dashed curves show the direct orbits, and the dotted curves represent the retrograde ones. The event horizon/singularity is denoted by a solid curve, while the horizontal dotted-dashed line at $2M$ represents the ergosphere in the equatorial plane.}
\label{fig1}
\end{figure}

\end{widetext}

For static black holes ($a_*=0$ and $\gamma=1$ on the top left hand plot),
the orbits with radii $r_{ms}$, $r_{mb}$ and $r_{ph}$, are located at $6M$, $4M$ and $3M$, respectively. As the spin increases from zero they pass through the ergosphere,
 and approach the even horizon $r_{+}$ at the maximal spin value $a_*=1$. For $\gamma<1$ (see the other three plots in Fig.~\ref{fig1}), there is a critical value $a_{*c}$ of the spin, where for each type of direct orbit the curves have a cut-off. The behavior of the retrograde orbits exhibits only a slight dependence on $\gamma$. The value of $a_{*c}$  is minimal for the radius $r_{ph}$, and maximal for $r_{ms}$, e.g., it is $a_{*c}\sim0.6$ and $a_{*c}\sim0.85$ for $\gamma=0.99$ (see the top right hand plot).  After reaching this critical value of the rotation, the marginally stable, marginally bound and photon orbits "jump" to the singularity, located at $r_s=2M/\gamma$. If the rotation exceeds the value $a_{*c}$, corresponding to the marginally stable orbit, then all the radii $r_{ms}$, $r_{mb}$, $r_{ph}$ and $r_{s}$ become degenerated. With increasing values of $\gamma$, these critical values are decreasing and even in the case of slow rotation we obtain at the singularity fully degenerated orbits.

\begin{widetext}

\begin{figure}
\centering
\includegraphics[width=.48\textwidth]{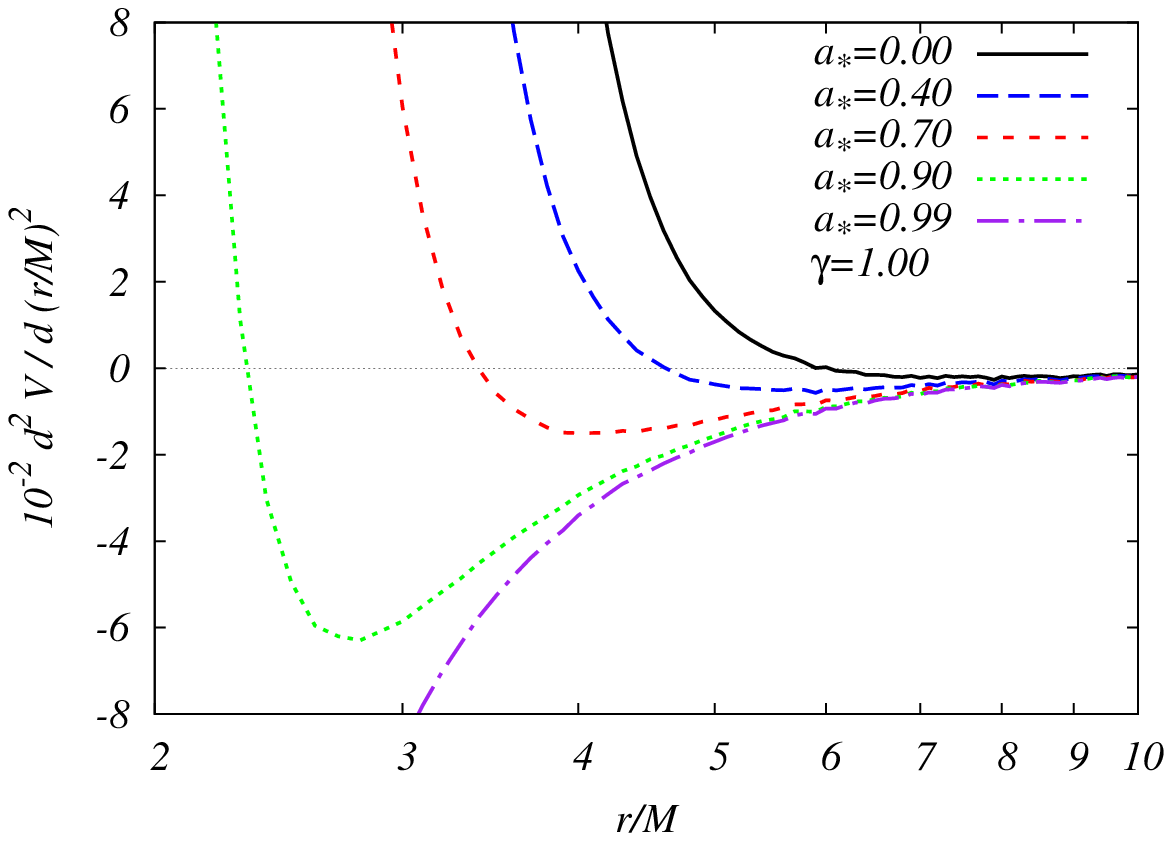}
\includegraphics[width=.48\textwidth]{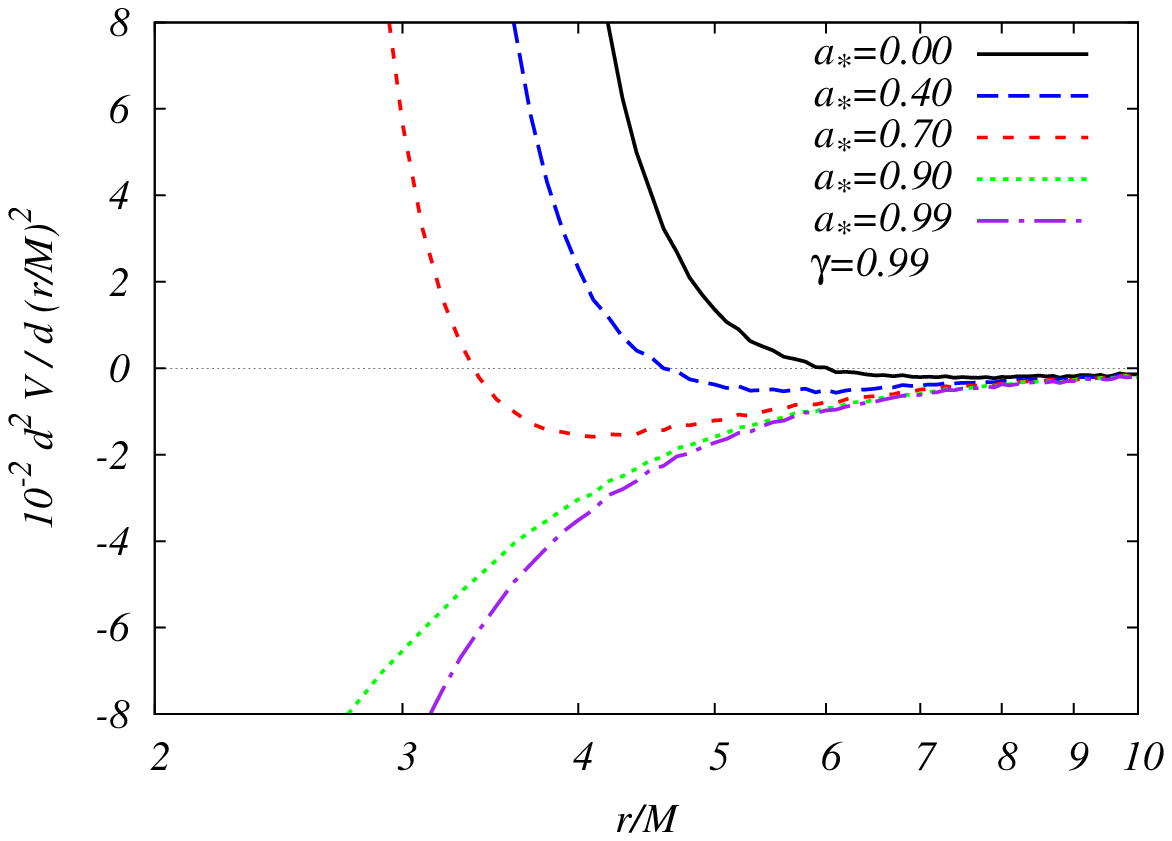}\\
\includegraphics[width=.48\textwidth]{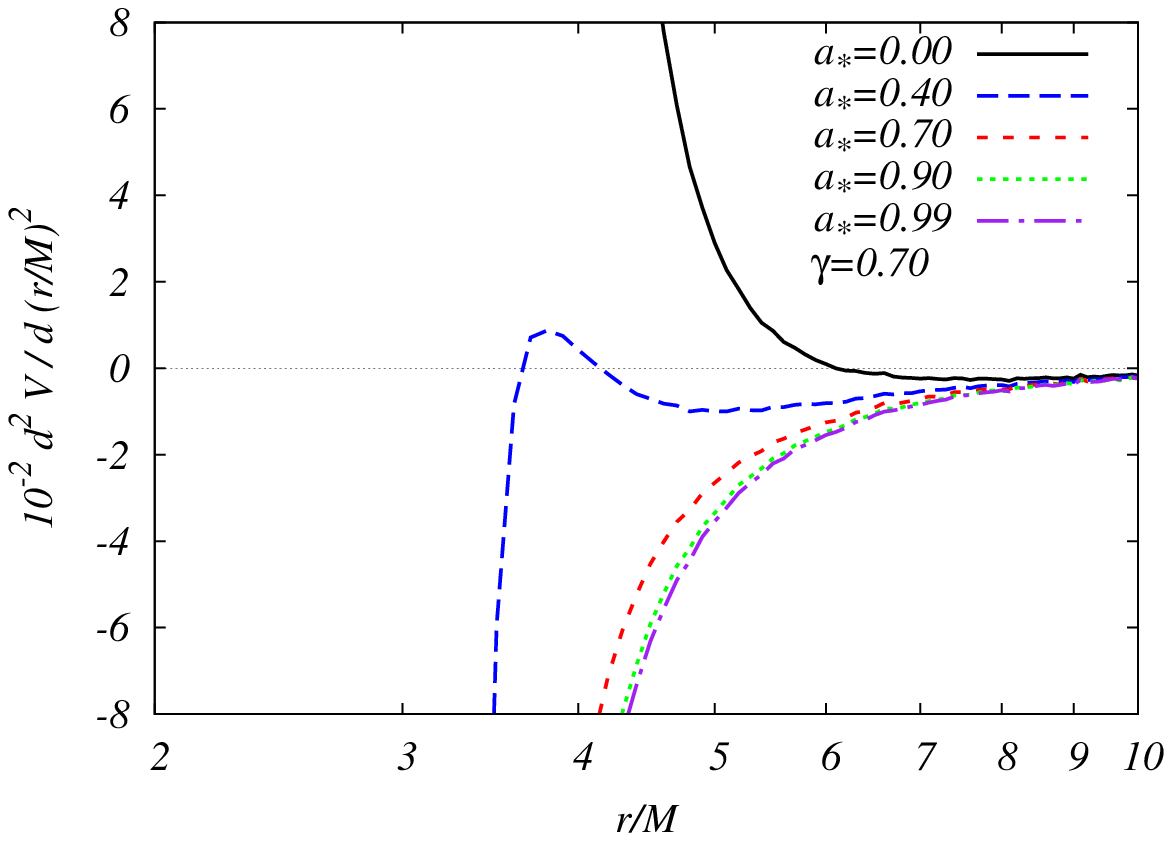}
\includegraphics[width=.48\textwidth]{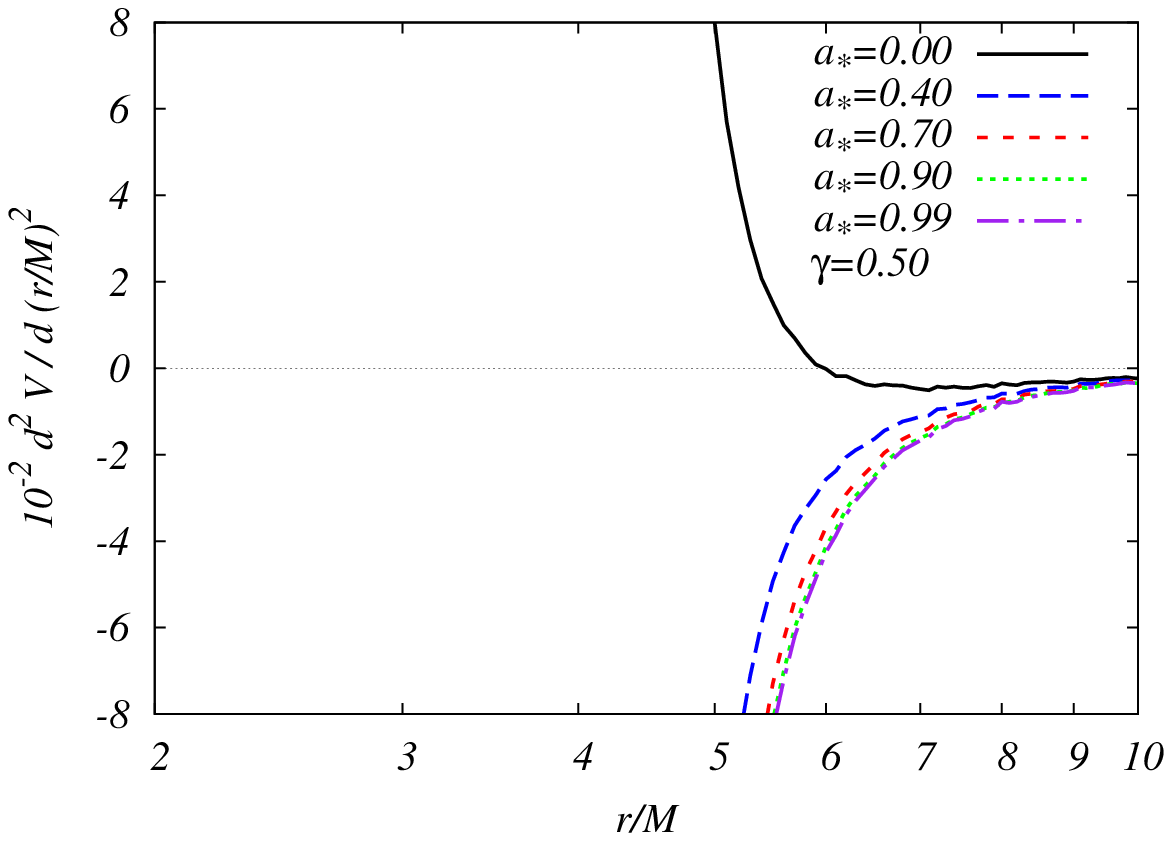}\\
\caption{The function $d^2V/d(r/M)^2$ versus $r/M$  for a rotating black hole, with ($\gamma=1$), and a naked singularity ($\gamma=0.99$, 0.7, and 0.5), with a total mass $M$. The spin parameter $a_*$ is set to 0.99, 0.9, 0.7 and 0, respectively.}
\label{fig2}
\end{figure}

\end{widetext}

As an illustration of the dependence of $r_{ms}$ on the spin $a_*$ and on the scalar charge parameter $\gamma$, we present in Fig.~\ref{fig2} the second order derivative of the effective potential with respect to the radial coordinate, given by Eq.~ (\ref{stable}), as a function of the radius $r$. Here the values of $\gamma$  are the same as in Fig.~(\ref{fig1}), and the spin parameter  $a_*$ takes different values. The top left hand panel shows the case of the black hole $(\gamma=1)$. For a static black hole $V_{,rr}$ vanishes at $6M$, and its zero shifts to lower radii with increasing spin parameter. For $a_*\sim0.9$,  $r_{ms}$ is already approaching the static limit at $r=2M$, and for $a_*\rightarrow1$, the marginally stable orbit enters the ergosphere, and it approaches the event horizon.
The other panels in Fig.~\ref{fig2}, presenting $V_{,rr}$ for the naked singularity ($\gamma<1$), show that Eq.~(\ref{stable}) has no longer solutions for higher values of the spin parameter, and $V_{,rr}$ remains negative everywhere. As a result, the particles rotating around the singularity have stable circular orbits in the whole spacetime, up to its boundary, i.e., to the singularity itself. With decreasing $\gamma$,  the critical value of $a_*$, above which $V_{,rr}$ has no zeros any more, is also decreasing. For $\gamma\lesssim 0.5$, only very slowly rotating naked singularities have a marginally stable orbit, which does not coincide with the singularity.

\section{The frame dragging effect}\label{mot1}

An interesting question related to the geodesic motion in the gravitational potential of the rotating naked singularity is the frame dragging effect, i.e., the relation between the rotational velocity $\omega$ of the spacetime itself, and the azimuthal component of the four-velocity of the freely falling (or rotating) massive test particles. The frame dragging frequency $\omega$ at the singularity can be calculated from Eq.~(\ref{omegaeq}) as
\begin{equation}
 \omega(r_s) =\lim_{x\rightarrow \sqrt{2/\gamma}} \frac{a_*}{2M\sqrt{2/\gamma^3}f^{1-\gamma}+2Ma_*^2}\:\label{Oms},
\end{equation}
where $x=\sqrt{r_s/M}=\sqrt{2/\gamma}$. From a physical point of view $\omega(r_s)$  can be interpreted as the angular velocity of the singularity, $\Omega^S=\omega(r_s)$. We note that the limit $\lim_{x\rightarrow \sqrt{2/\gamma}} f^{1-\gamma}$ vanishes for $\gamma<1$, but it becomes unity for $\gamma =1$. The latter result is not trivial, but if it holds, we reobtain $\omega(r_e)$ for the Kerr solution, since $r_s\rightarrow r_e$ for $\gamma=1$. For the rotating naked singularity metric given by Eq.~(\ref{ds2}), the expression (\ref{Oms}) reduces to
\[
 \Omega^S=\omega(r_s) = \frac{1}{2Ma_*},
\]
which is clearly valid only for a rotating singularity, since $\Omega^S$ would diverge in the static case. Even if $a_*\ne0$ holds, the explicit expression for $\Omega^S$ might be puzzling, since it does not depend on $\gamma$, and shows that the angular velocity of a rotating naked singularity is inversely proportional to its spin parameter.
We would expect that the angular frequency of any rotating object to be proportional to the angular momentum of the body - and this is indeed true almost over the whole spacetime, except in a very small domain, close to the naked singularity. In a very close vicinity of the naked singularity, at $r_s$, $\omega $ becomes bigger  for lower values of $a_*$, as compared to the values corresponding to  higher spins.
We can determine the radius $r_{i}$ at which the proportionality relation $\omega(x,a_{*1})>\omega(x,a_{*2})$ for $a_{*1}>a_{*2}$ is inverted by inserting Eq.~(\ref{omegaeq}) into the equation $\omega(x_{i},a_{*1})=\omega(x_{i},a_{*2})$,  with $x_i=\sqrt{r_i/M}$. The result is $a_{*1}{\mathscr A}(x_i,a_{*2})=a_{*2}{\mathscr A}(x_i,a_{*1})$, which can be further simplified to the equation
\begin{equation}
x^4_if^{1-\gamma}=a_{*1}a_{*2}(2-f^{\gamma})\;.\label{ri}
\end{equation}

For $\gamma=1$ this equation reduces to the cubic equation $x_i^4-a_{*1}a_{*2}x_i^2-a_{*1}a_{*2}=0$ , which has the triple real root $x_i=0$, showing that, as expected,  there is no inversion in the $\omega-a_*$ relation for Kerr black holes. For $\gamma<1$,  the biggest real root of Eq.~(\ref{ri}) provides the outer boundary of the region between $r_s$ and $r_i$,  where the frame dragging effect of the spacetime  with the spin parameter $a_{1*}$ is weaker than the frame dragging effect measured for a lower spin value $a_{2*}$. Outside this region, i.e., for all  radii $x$ greater than $x_{i}$, the relation $\omega(x,a_{*1})>\omega(x,a_{*2})$ holds.  Fig.~\ref{fig3}, where we plotted $\omega $ as a function of the radial distance for naked singularities with same mass and same scalar charge ($\gamma=0.9$), but different spin parameters, shows this behavior. In Fig.~\ref{fig3}, the solid black line, representing the frame dragging frequency obtained for the naked singularity with the spin value of 0.99, has the highest values at radii $r/M$ greater than $\sim r_s/M+10^{-4}/\gamma$.  For lower radii, a radius $r_i$, where the $\omega $ curves for lower spin values intersect the black solid curve, always exists. The lower the spin value is, the closer $r_i$ shifts to $r_s$. For radii less than $r_i$, the frame dragging frequency for $a_*<0.99$ becomes higher  than the value obtained for $a_*=0.99$. This result holds for any pair of spin values, i.e., there exist radii $r_i$, where for any value of $a_{*1}$, the curve $\omega $, and the curve for $a_{*1}>a_{*2}$, intersect each other. In all cases we obtain the relation  $\omega(x,a_{1*})<\omega(x,a_{2*})$ for $x<\sqrt{r_i/M}$.
\begin{figure}
\centering
\includegraphics[width=.48\textwidth]{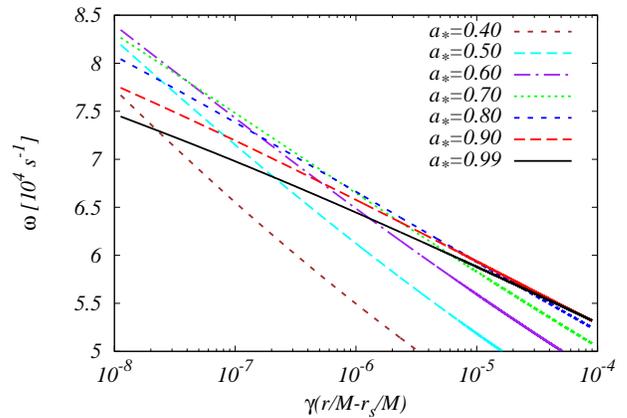}
\caption{The radial profile of $\omega$ near the rotating naked singularity with $\gamma=0.9$, for different spin parameters.}
\label{fig3}
\end{figure}

A similar pathological behavior can be found for the angular velocity $\Omega$ of massive test particles.
 For the first term of $\mathscr{B}$ appearing in Eq.~(\ref{Omega2}), as we approach the singularity we obtain the limit
\begin{eqnarray}\lim_{x\rightarrow\sqrt{2/\gamma}}f^{(1-\gamma)}h&=& \lim_{x\rightarrow\sqrt{2/\gamma}}f^{1-\gamma}\left(1-\frac{1-\gamma}{\gamma x^2f} \right)^{1/2}\nonumber\\
&=& \frac{1}{2}\sqrt{\gamma(1-\gamma)}\lim_{x\rightarrow\sqrt{2/\gamma}}f^{1/2-\gamma}.
\end{eqnarray}
 For the angular velocity we obtain
 \begin{equation}
\lim_{x\rightarrow\sqrt{2/\gamma}}\Omega=M^{-1}\begin{cases}
0\:, & 1/2<\gamma<1\:,\\
(2+a_*)^{-1}\:, & \gamma=1/2\:,\\
a_{*}^{-1}\:, & 0<\gamma<1/2\end{cases}\;\label{Olim}.
\end{equation}
By supposing that the naked singularity rotates fast enough, so that $r_{ph}=r_{s}$, then, if $1/2<\gamma <1$, for the orbits located close to the singularity $\Omega$ decreases to zero. For $\gamma=1/2$, the angular velocity of the rotating particles has a non-vanishing value at the singularity, and it remains finite even in the static case, ($\lim_{x\rightarrow\sqrt{2/\gamma}}f^{1/2-\gamma}=1$ is supposed again). For $\gamma<1/2$, $\Omega$ is proportional to $a^{-1}_*$, and for $a_*\rightarrow 0$ it diverges to infinity, as $\omega$ also does. Since at the singularity $\Omega_{max}(r_s)=2\omega(r_s)$ coincides with $\Omega$, the difference $\Omega -\omega$ also diverges at $r_s$ for a vanishing spin parameter. However, this divergence does not involve any violation of causality, since the velocity of the orbiting particles, measured in the locally non rotating frame, does not exceed the speed of light, i.e., $|v|\le1$ still holds, where
\[
|v|=\frac{g_{\phi\phi}}{\sqrt{g^2_{t\phi}-g_{tt}g_{\phi\phi}}}|\Omega-\omega|,
\]
for circular equatorial orbits. At $r_s$ we have $g_{tt}=0$, and $|v|=|\Omega/\omega-1|=1$, a relation that provides the upper limit of the velocity for time-like particles that satisfy the condition (\ref{cond}) when moving along circular geodesics in the equatorial plane.

In Fig.~\ref{fig4} we have plotted $\Omega$, $\Omega_{min}$, $\Omega_{max}$ and $\omega$ for the same values of $a_*$ and $\gamma$ as we have used in Fig.~\ref{fig2}. The top left hand panels show the frame dragging effect in the Kerr spacetime. When $\omega=0$ (the static case), for $r>r_{ph}=3M$ the orbiting particles can have a Keplerian rotation, with the frequency $\Omega$. At lower radii there are no circular orbits, and the particles fall freely onto the event horizon, with a velocity with azimuthal component restricted to values between $\Omega_{min}$ and $\Omega_{max}$. For the static black hole at the even horizon these values tend to $\omega=0$, i.e., there is no frame dragging. With increasing $a_*$, the frame dragging effect becomes more and more important: as the orbital radius is approaching $r_{+}$, the particle is dragged with the rotating spacetime. No static observer exists in the ergosphere, since any particle passing through the static limit $(g_{tt}=0)$ is forced to rotate with a positive angular velocity. For rapid rotations $a_*\rightarrow1$, $r_{ph}$ is also approaching the event horizon, where the particles move along circular orbits with the angular velocity of the horizon, namely, $\Omega^H=\omega(r_{+})$. At $r_{+}$ we obtain $\Omega=\Omega_{min}=\Omega_{max}=\omega=\Omega^H$, as $a_*\rightarrow 1$.

\begin{widetext}

\begin{figure}
\centering
\includegraphics[width=.48\textwidth]{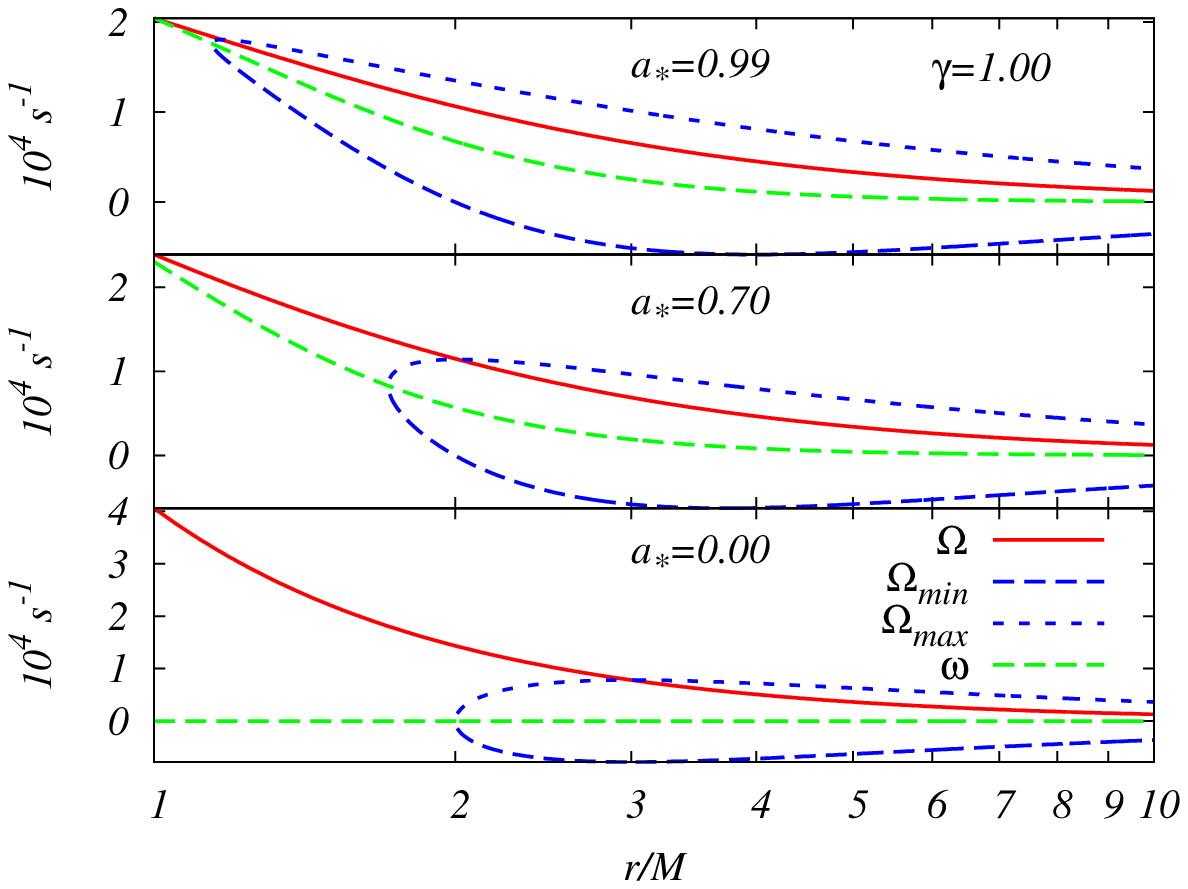}
\includegraphics[width=.48\textwidth]{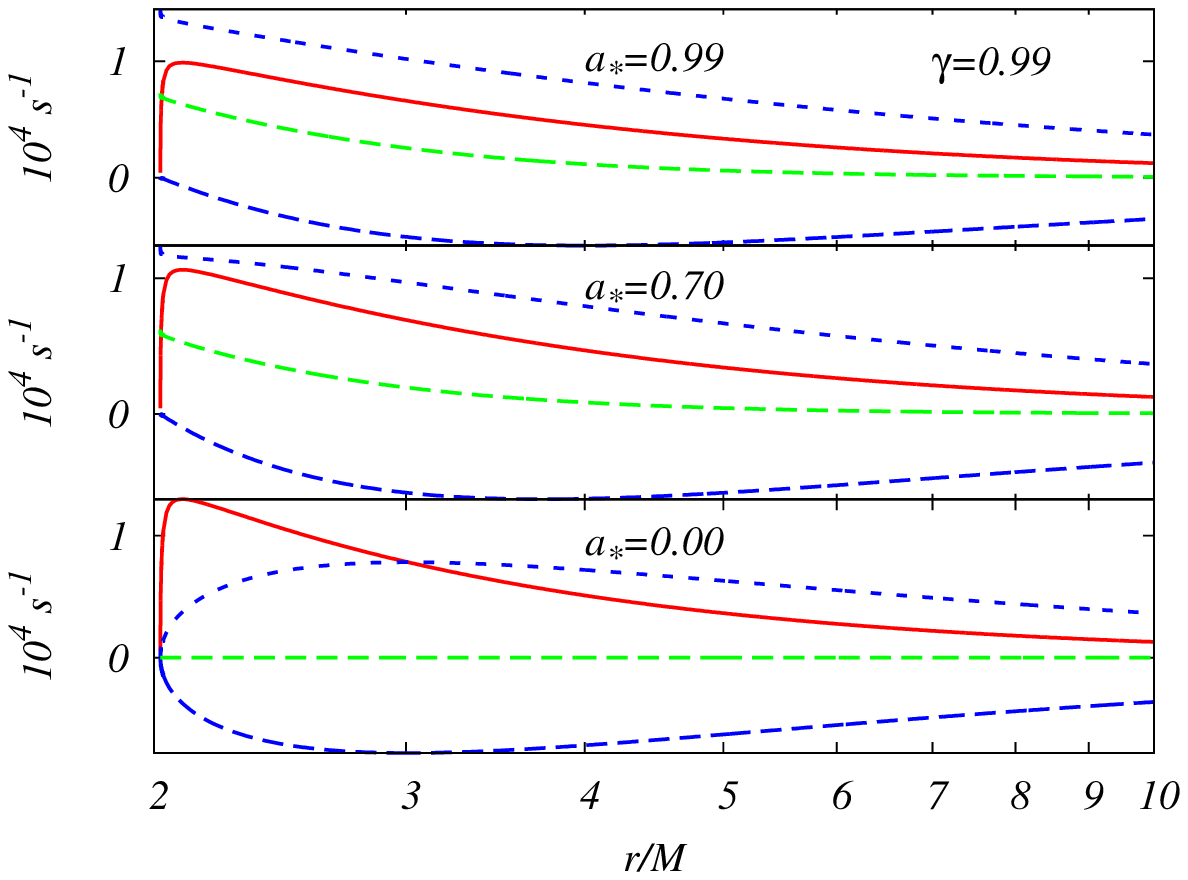}\\
\includegraphics[width=.48\textwidth]{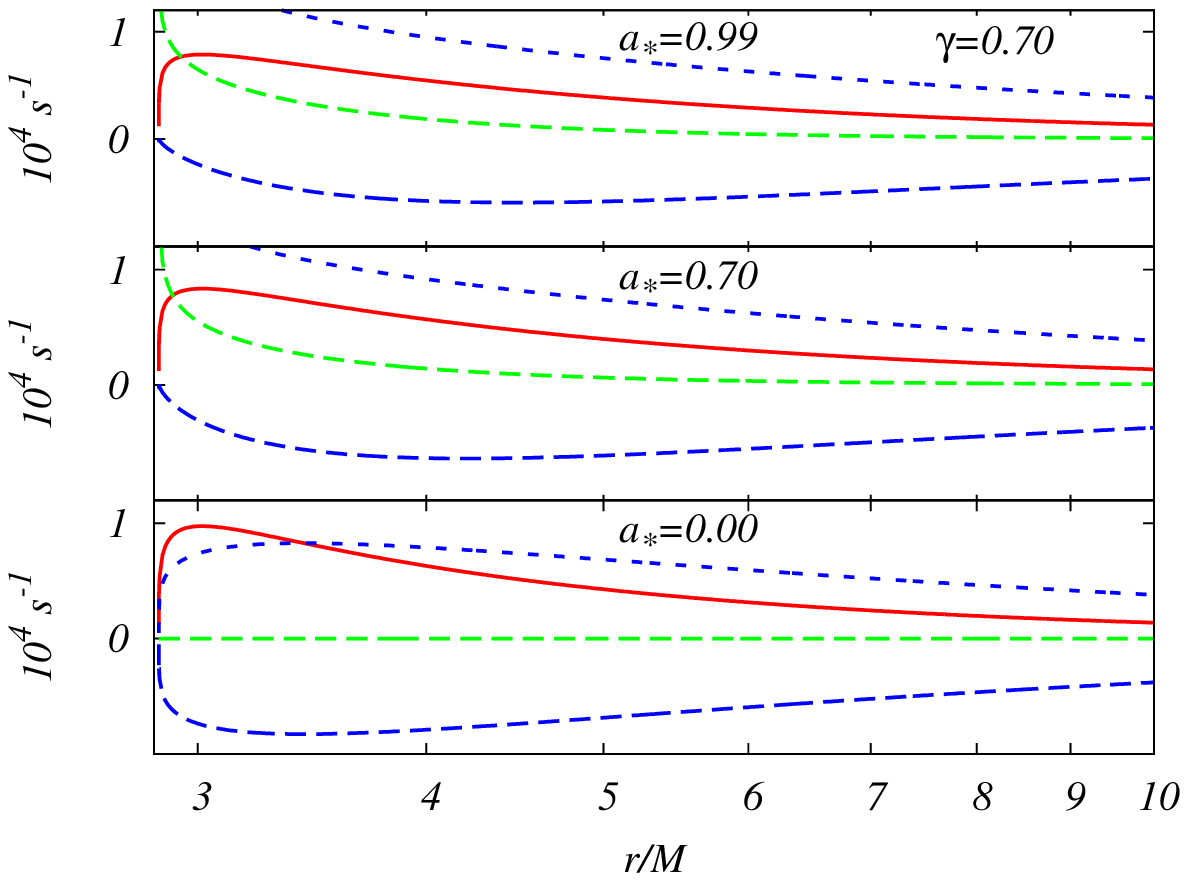}
\includegraphics[width=.48\textwidth]{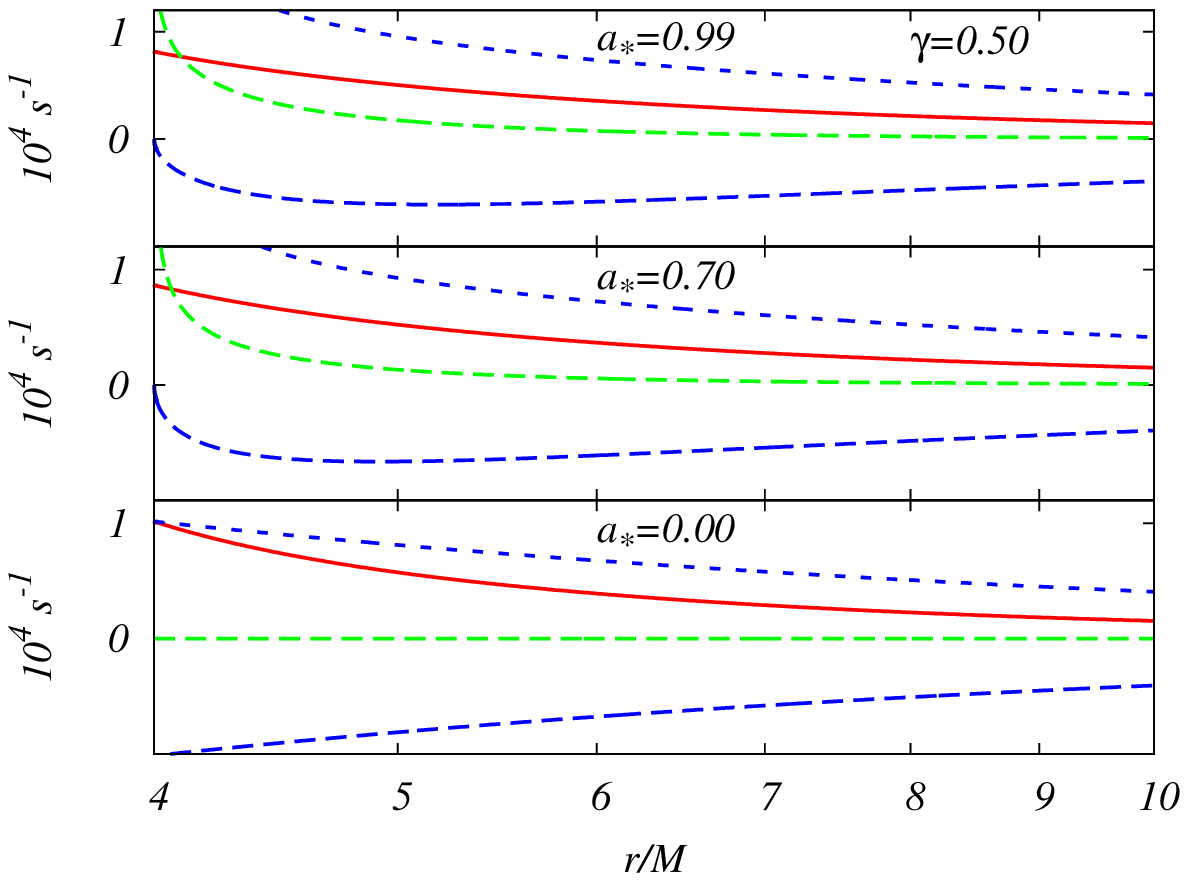}\\
\includegraphics[width=.48\textwidth]{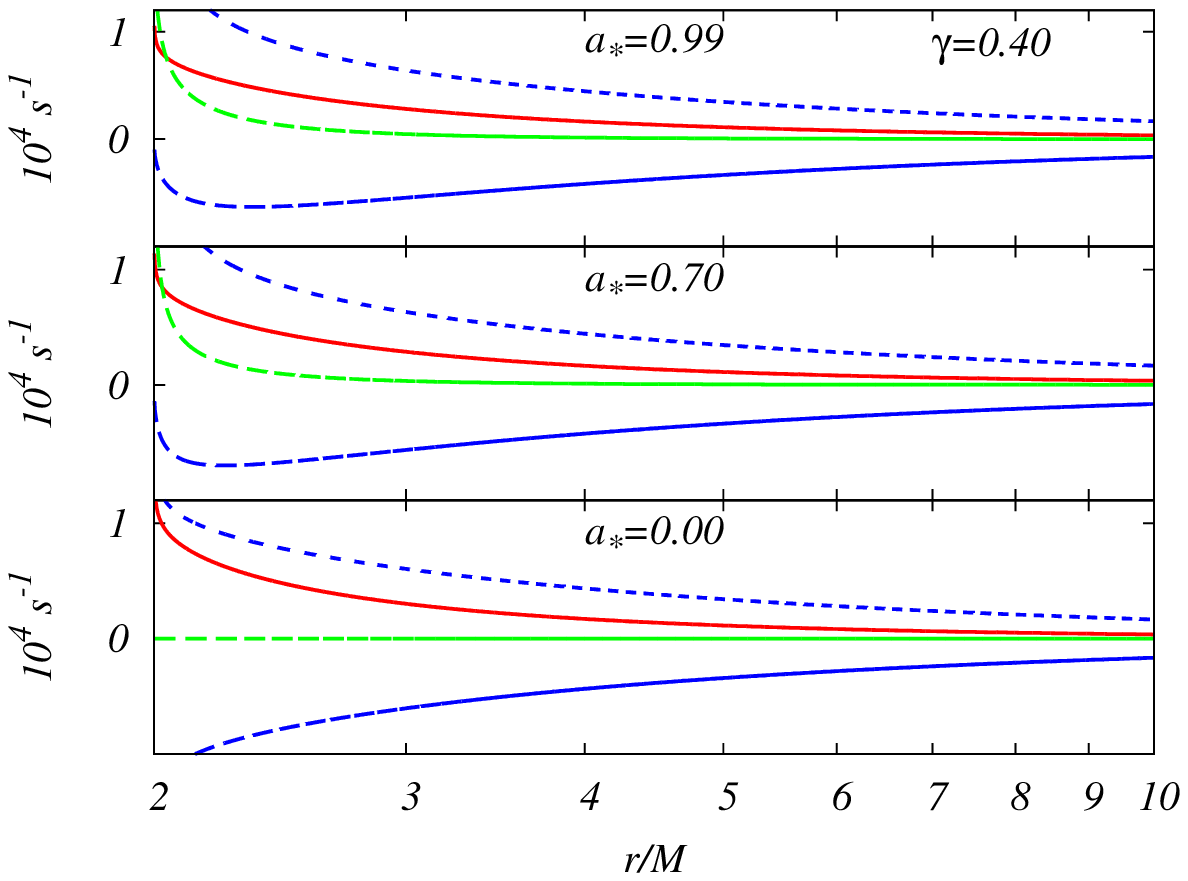}
\includegraphics[width=.48\textwidth]{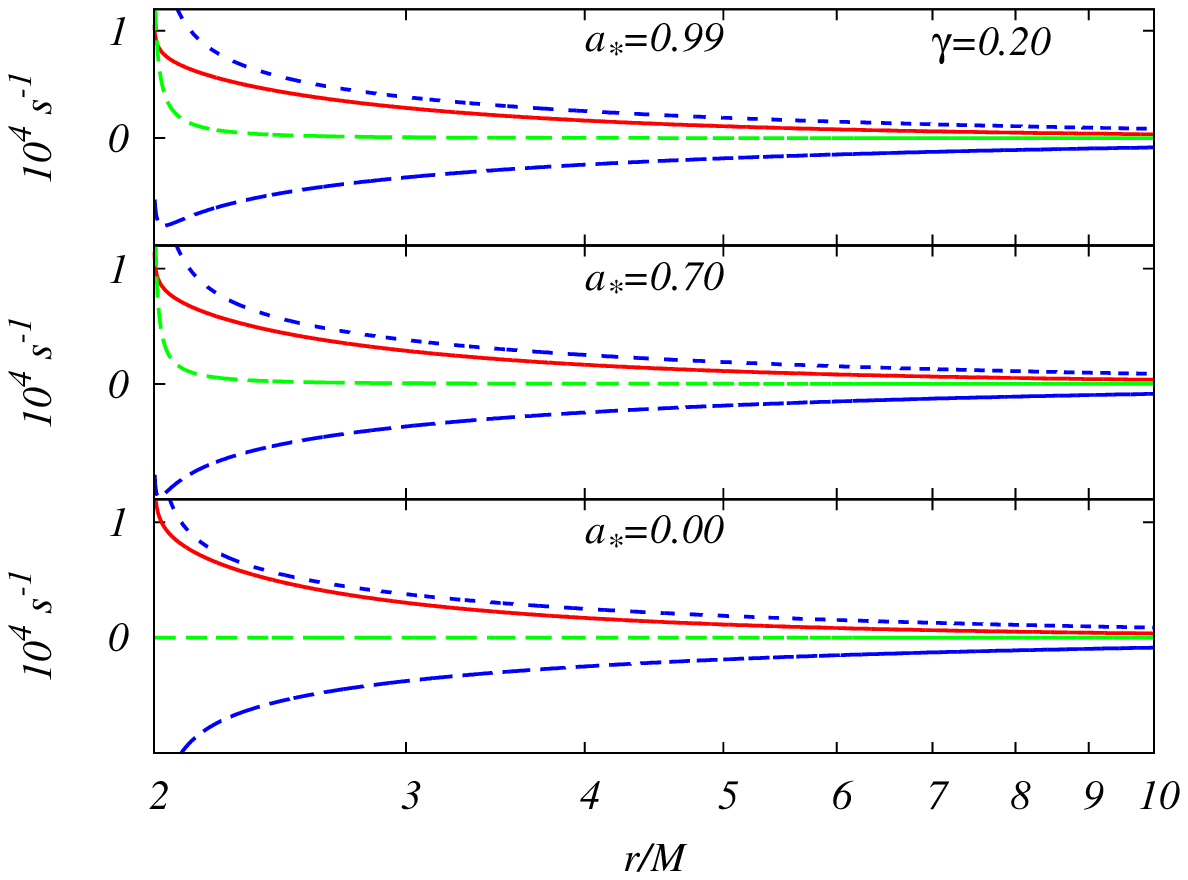}\\
\caption{The radial profiles of $\Omega$, $\Omega_{min}$, $\Omega_{max}$ and $\omega$ for black holes with $\gamma=1$ and for the rotating naked singularity with $\gamma=0.99$, 0.9, 0.7, 0.5, 0.4, and 0.2, respectively. The spin parameter $a_*$ is set to 0.99, 0.7 and 0, respectively.}
\label{fig4}
\end{figure}

\end{widetext}

These effects are considerably different for rotating naked singularities, presented in the rest of the panels in Fig.~{\ref{fig3}}.
At $r=r_{s}$, the lapse function for the metric vanishes, $g_{tt}=0$, as it does in the static limit for the black holes, and Eqs.~(\ref{Omin}) and (\ref{Omax}) give $\Omega_{min}=0$ and $\Omega_{max}=2\omega=1/Ma_*$, without fixing the value of $\Omega$ at the singularity. The result that at the singularity, depending on the values of $\gamma$, $\Omega$ is either zero, or takes some non-zero values between 0 and $2\omega$,  follows from Eq.~(\ref{Olim}).
For relatively small scalar charges, $\gamma=0.99$, the static case is still similar to the static case of the black holes: the marginally stable orbit is located at around $6M$ (the top right hand panel in Fig.~\ref{fig2}), and for the orbits with radii greater than $r_{ph}\sim 3M$, the Keplerian rotation can still be maintained. At lower radii $\Omega$ is restricted to values between $\Omega_{min}$ and $\Omega_{max}$, and $\Omega$ vanishes at the singularity (as it does at the event horizon of a Schwarzschild black hole). For rotating singularities with $\gamma=0.99$ and $a_*=0.7$, $r_{ms}$ is still greater than $4M$, but the photon radius has jumped to the singularity (the top right hand panel in Fig.~\ref{fig1}). Particles can therefore move along circular orbits in the whole spacetime, although not all these orbits are stable. However, at the singularity, located at $\sim2M$, $\Omega$ vanishes, even if $\omega$ has a finite, non-zero value. For $\gamma=1/2$, $\Omega$ has already a non-zero value at the singularity, which is inversely proportional to $2+a_*$, but it is still less than $\omega$, as shown in the middle right hand panel in Fig ~\ref{fig4}. The panels in the bottom of Fig.~\ref{fig4} show that both $\Omega$ and $\omega$ increase very rapidly close to the singularity - in fact $\Omega$ exceeds the frame dragging frequency at $r_s$ as  $\Omega(r_s)\rightarrow1/Ma_*$, and the relation $\omega(r_s)\rightarrow1/2Ma_*$ holds for $\gamma<1/2$. The relation $\Omega>\omega$ for $r\rightarrow r_s$ is shown in Fig.~\ref{fig5}, where we have plotted the ratio $\Omega/\omega$ as a function of the radius. As mentioned before, for $a_*\rightarrow0$ this ratio tends to $1/2$ at the singularity, the particles orbiting at $r_s$ have finite velocities, with the speed of the light like particles tending towards the speed of light.

\begin{widetext}

\begin{figure}
\centering
\includegraphics[width=.48\textwidth]{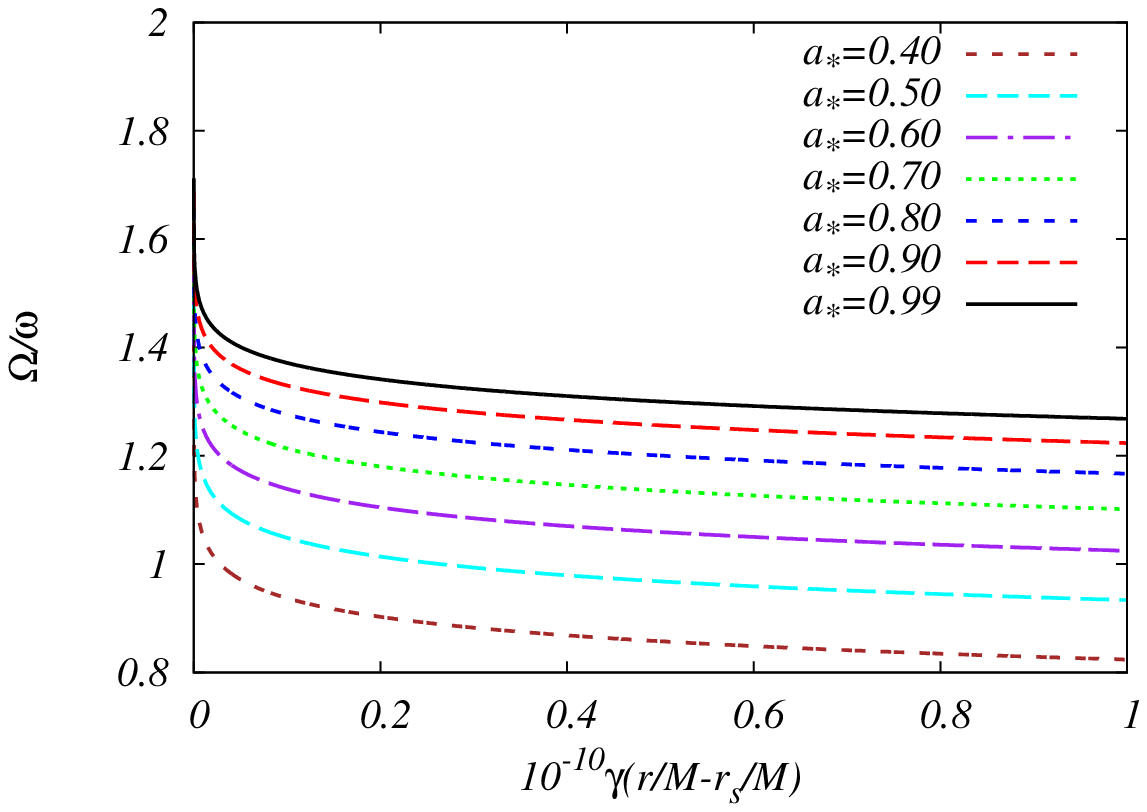}
\includegraphics[width=.48\textwidth]{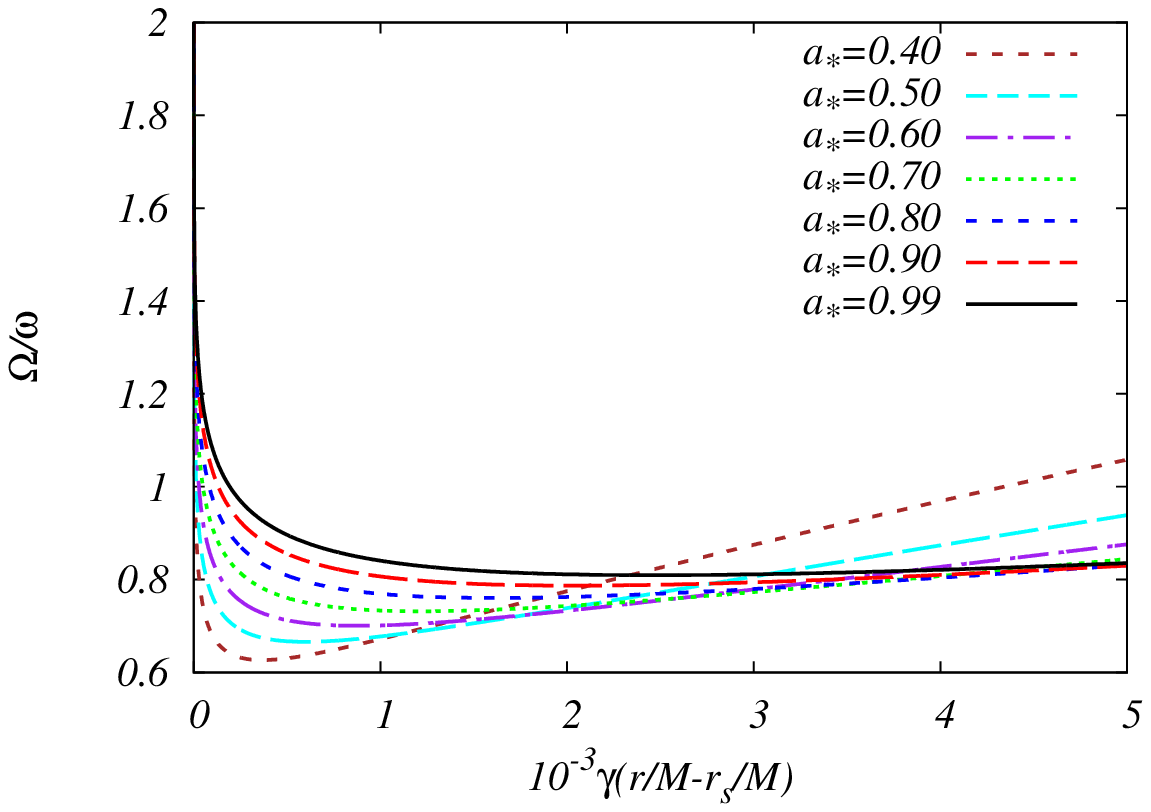}\\
\caption{The radial profiles of the ratio $\Omega /\omega$ for $\gamma=0.4$ (left hand panel) and $\gamma=0.2$ (right hand panel), for different values of the spin parameter.}
\label{fig5}
\end{figure}

\end{widetext}

The angular velocity of particles moving along stable circular geodesics around black holes is maximal at the marginally stable orbit, and decreases monotonically, with the increasing coordinate radius $r$. Eq.~(\ref{Olim}), and Fig.~\ref{fig4}, show that this is not the case for the particles rotating in the gravitational potential of rotating naked singularities with $\gamma>1/2$. At the singularity $\Omega$ drops to zero from its maximal value at $r_{max}$, which is still close to the singularity ($r_{max}-r_s\ll M$). By inserting Eq.~(\ref{Omega2}) into the condition $\Omega_{,r}=0$ we obtain the equation
\begin{equation}
3\gamma^2x^4-6\gamma(\gamma+1)x^2+4\gamma^2+6\gamma+2=0,\label{Omega_r}
\end{equation}
with the solution
\begin{equation}
x^2_{max,\pm}=\frac{1+\gamma\pm\sqrt{( 1 -\gamma^2 )/3}}{\gamma}.\label{x2pm}
\end{equation}
For the limiting case of the Kerr black holes ($\gamma=1$),  Eq.~(\ref{x2pm}) has  only one  root, $x^2_{max,+}=x^2_{max,-}=2$, or $r_{max}=2M$, which is  due to  multiplication of $\Omega_{,r}$ by $f(r)$ in the computation of Eq.~(\ref{Omega_r}). At $r=2M$, the derivative of $\Omega$ is not zero, indicating a monotonic decrease in the angular velocity with increasing radial coordinate.
For $\gamma\ne 1$ Eq.~(\ref{x2pm}) has two roots, where $x^2_{max,+}=r_{max}/M$ is the value of the radial coordinate where $\Omega$ is maximal.
Since Eq.~(\ref{x2pm}) does not depend on the spin parameter $a_*$, the dimensionless radius $x_{max,+}$ is only determined by the parameter $\gamma$ of the singularity. The location of the marginally stable orbit $r_{ms}$ still depends on $a_*$, as in the case of Kerr black holes, and it decreases as the singularity rotates faster and faster.
At orbits with radii higher than $r_{max}$, the angular velocity is increasing with decreasing $r$,  but if $r_{max}$ is greater than $r_{ms}$, then there is an annulus between the inner edge of the accretion disk and $r_{max}$, where for particles moving along circular orbits $\Omega$ is decreasing for small distances from the singularity.

\section{Standard accretion disks}\label{std}

 Accretion discs are flattened astronomical objects made of rapidly rotating gas which slowly spirals onto a central gravitating body, with its gravitational energy degraded to heat. A fraction of the heat converts into
radiation, which partially escapes, and cools down the accretion disc.
The only information that we have about accretion disk physics comes from
this radiation, when it reaches radio, optical and $X$-ray telescopes,
allowing astronomers to analyze its electromagnetic spectrum, and its time
variability. The efficient cooling via the radiation over the disk surface prevents the disk
from cumulating the heat generated by stresses and dynamical
friction. In turn, this equilibrium causes the disk to stabilize
its thin vertical size. The thin disk has an inner edge at the
marginally stable orbit of the compact object potential, and the
accreting plasma has a Keplerian motion in higher orbits.

For the general relativistic case the theory of mass accretion around rotating black holes was developed by Novikov and Thorne \cite{NoTh73}. They
extended the steady-state thin disk models introduced in
\cite{ShSu73} to the case of the curved space-times, by adopting the equatorial
approximation for the stationary and axisymmetric geometry.
For a steady-state thin accretion disk described in the
cylindrical coordinate system $(r,\phi ,z=\cos\theta)$ most of the
matter lies close to the radial plane.
Hence its vertical size (defined along the $z$-axis)
is negligible, as compared to its horizontal extension (defined along the
radial direction $r$), i.e, the disk height $H$, equal to the maximum
half thickness of the disk, is always much smaller than the characteristic radius $R$ of
the disk, $H \ll R$. The thin disk is in hydrodynamical equilibrium, and the pressure gradient
and a vertical entropy gradient in the accreting matter are negligible.
In the steady-state accretion disk models, the mass accretion rate $\dot{M%
}_{0} $ is supposed to be constant in time, and the physical quantities of
the accreting matter are averaged over a characteristic time scale, e.g. $%
\Delta t$, and over the azimuthal angle $\Delta \phi =2\pi $, for a total period of
the orbits and for the height $H$. The plasma moves in Keplerian orbits around
the compact object, with a rotational velocity $\Omega $, and the plasma
particles have a specific energy $\widetilde{E}$, and specific angular
momentum $\widetilde{L}$, which depend only on the radii of the orbits. The
particles are orbiting with the four-velocity $u^{\mu }$ in a disk having an
averaged surface density $\Sigma $. The accreting matter is modeled by an
anisotropic fluid source, where the rest mass density $\rho _{0}$ (the specific internal energy
is neglected), the energy flow vector $q^{\mu }$ and the stress tensor $%
t^{\mu \nu }$ are measured in the averaged rest-frame. The energy-momentum
tensor describing this source takes the form
\begin{equation}
T^{\mu \nu }=\rho _{0}u^{\mu }u^{\nu }+2u^{(\mu }q^{\nu )}+t^{\mu \nu }\;,
\end{equation}
where $u_{\mu }q^{\mu }=0$, $u_{\mu }t^{\mu \nu }=0$. The four-vectors of
the energy and of the angular momentum flux are defined by
$-E^{\mu }\equiv T_{{}}^{\mu }{}_{\nu }(\partial /\partial t)^{\nu }$ and $
J^{\mu }\equiv T_{{}}^{\mu }{}_{\nu }(\partial /\partial \phi )^{\nu }$,
respectively. The four dimensional conservation laws
of the rest mass, of the energy and of the angular momentum of the plasma
provide the structure equations of the thin disk. From the structure equations
the flux of the radiant energy over the disk can be expressed as
\citep{PaTh74, Th74}
\begin{equation}
F(r)=-\frac{\dot{M}_0}{4\pi\sqrt{-g}} \frac{\Omega_{,r}}{(\widetilde{E}%
-\Omega\widetilde{L})^{2}} \int_{r_{ms}}^{r}(\widetilde{E}-\Omega\widetilde{L%
})\widetilde{L}_{,r}dr\;,  \label{F}
\end{equation}
where the no-torque inner boundary conditions were prescribed \cite{PaTh74}. This means that the torque vanishes at the inner edge of the disk,
since the matter at the marginally stable orbit $r_{ms}$ falls
freely into the black hole, and cannot exert considerable torque on the
disk. The latter assumption is valid as long as strong magnetic fields do
not exist in the plunging region, where matter falls into the hole.

The geometry of the space-time near to the equator, or the metric potential determines the radial dependence of $\Omega $, $\widetilde{E}$ and
 $\widetilde{L}$ for the particles moving on
circular orbits around the central object. We can therefore calculate the
averaged radial distribution of photon emission for accretion disks around
the rotating singularity in the equatorial approximation, by applying the
flux integral Eq.~(\ref{F}). Evaluating of the specific energy at the inner edge of the disk, we can also determine the efficiency of conversion of the rest mass into outgoing radiation.

The accreting matter in the steady-state thin disk model is supposed to be
in thermodynamical equilibrium. Then the radiation emitted by the disk
surface can be considered as a perfect black body radiation, where the
energy flux is given by $F(r)=\sigma _{SB}T^{4}(r)$ ($\sigma _{SB}$ is the
Stefan-Boltzmann constant) and the observed luminosity $L\left( \nu \right) $ has a redshifted black body spectrum \cite{To02}:
\begin{equation}
L\left( \nu \right) =4\pi d^{2}I\left( \nu \right) =\frac{8}{\pi c^2 }\cos i \int_{r_{i}}^{r_{f}}\int_0^{2\pi}\frac{\nu^{3}_e r d\phi dr }{\exp \left( \nu_e/T\right) -1}.\label{L}
\end{equation}
Here $d$ is the distance to the source, $I(\nu )$ is the Planck
distribution function, $i $ is the disk inclination angle (we set it to zero), and $r_{i}$ and $r_{f}$ indicate the position of the inner and outer edge of the disk,
respectively. We take $r_{i}=r_{ms}$ and $r_{f}\rightarrow \infty $, since
we expect the flux over the disk surface vanishes at $r\rightarrow \infty $
for any kind of general relativistic compact object geometry. The emitted frequency is given by $\nu_e=\nu(1+z)$, where the redshift factor can be written as
\begin{equation}
1+z=\frac{1+\Omega r \sin \phi \sin \gamma }{\sqrt{ -g_{tt} - 2 \Omega g_{t\phi} - \Omega^2 g_{\phi\phi}}}\;,
\end{equation}
where we have neglected the light bending \cite{Lu79,BMT01}.

The efficiency $\epsilon $ with which the central object converts rest mass into outgoing radiation is the other important physical parameter characterizing the properties of the accretion disks. The efficiency
is defined by the ratio of two rates measured at infinity: the rate of the
radiation of the energy of the photons escaping from the disk surface to infinity, and the rate at which mass-energy is transported to the compact object. If all the emitted photons can escape to infinity, the efficiency depends only on the specific energy measured at the marginally stable orbit $r_{ms}$,
\begin{equation}
\epsilon = 1 - \left.\widetilde{E}\right|_{r=r_{ms}}\;.  \label{epsilon}
\end{equation}
For Schwarzschild black holes the efficiency is about 6\%, no matter if we
consider the photon capture by the black hole, or not. Ignoring the capture
of radiation by the black hole, $\epsilon$ is found to be 42\% for rapidly
rotating black holes, whereas the efficiency is 40\% with photon capture in
the Kerr potential.

\section{Electromagnetic signatures of accretion disks around rotating naked singularities}\label{el}

After analyzing, in Sections \ref{mot} and \ref{mot1}, the circular geodesic motion around a rotating naked singularity, we are now ready to discuss the properties of the disk radiation for standard accretion disk models in the spacetime of the naked singularity. In Fig.~\ref{fig6} we present the flux profile, calculated from Eq.~(\ref{F}), for the physical parameters of the configurations already shown in Fig.~\ref{fig2}. In the following we set the total mass to $5M_{\odot}$, and the accretion rate to $10^{-12}M_{\odot}$/yr.
The top left hand panel shows $F(r)$ for the static and the rotating black holes. The dependence of the flux distribution on the spin has distinct features: the inner edge of the disk is located at  $r_{ms}=6M$ for the static black hole, and shifts to lower radii, approaching $M$, as the black hole spins up to $a_*=1$. The radii of the marginally stable orbits are determined by the zeros of $V_{,rr}$ in Fig.~\ref{fig2}. With increasing spin, the maximal flux is also increasing with at least three orders of magnitude, as compared to the cases with $a_*=0$
and $a_*=0.99$. For higher values of the spin, the locations of the maxima of the spectra also shift to lower radii, located closer to the inner edge of the disk.

\begin{widetext}

\begin{figure}
\centering
\includegraphics[width=.48\textwidth]{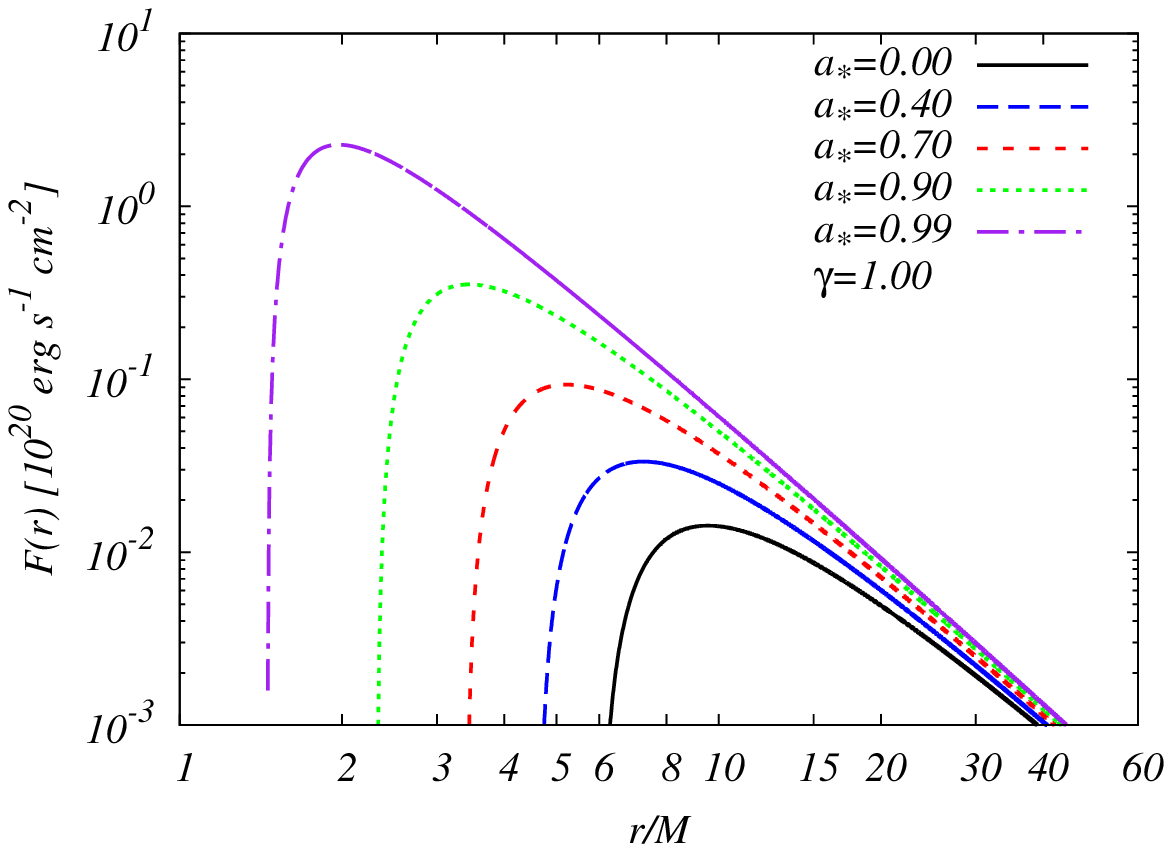}
\includegraphics[width=.48\textwidth]{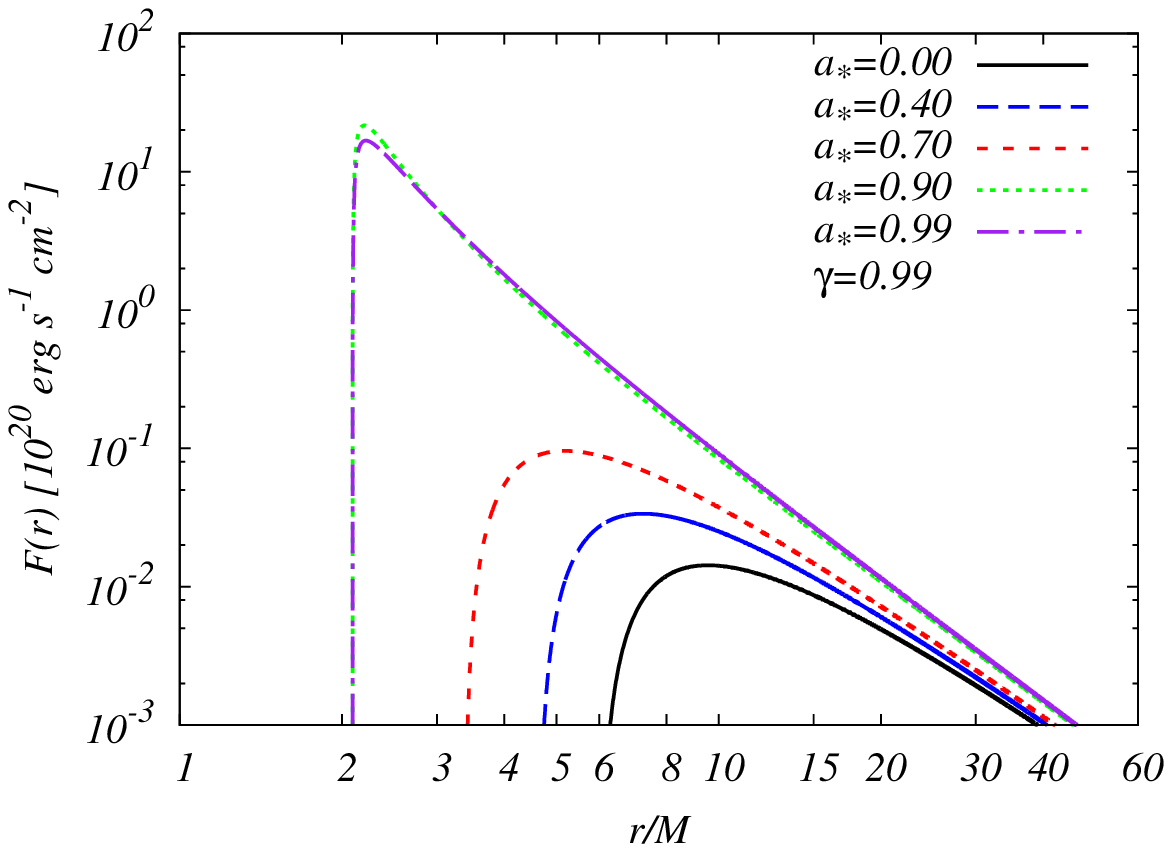}\\
\includegraphics[width=.48\textwidth]{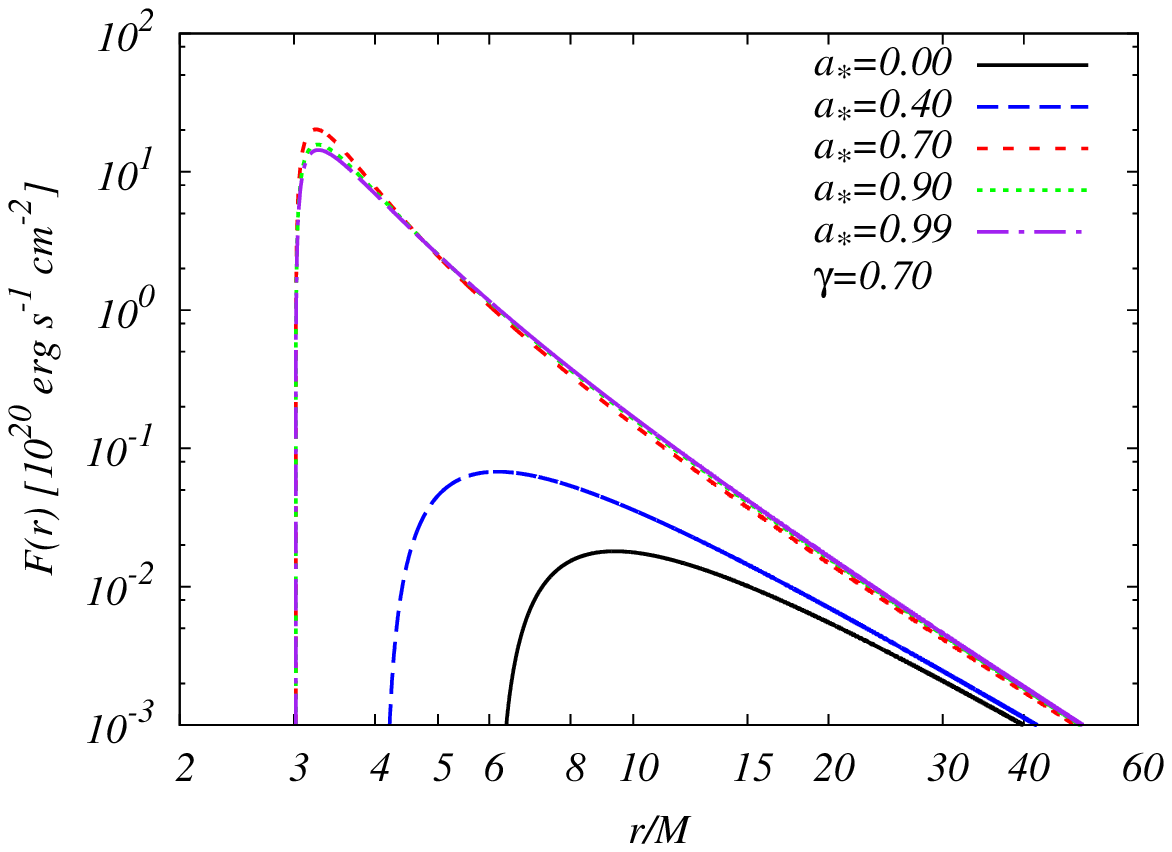}
\includegraphics[width=.48\textwidth]{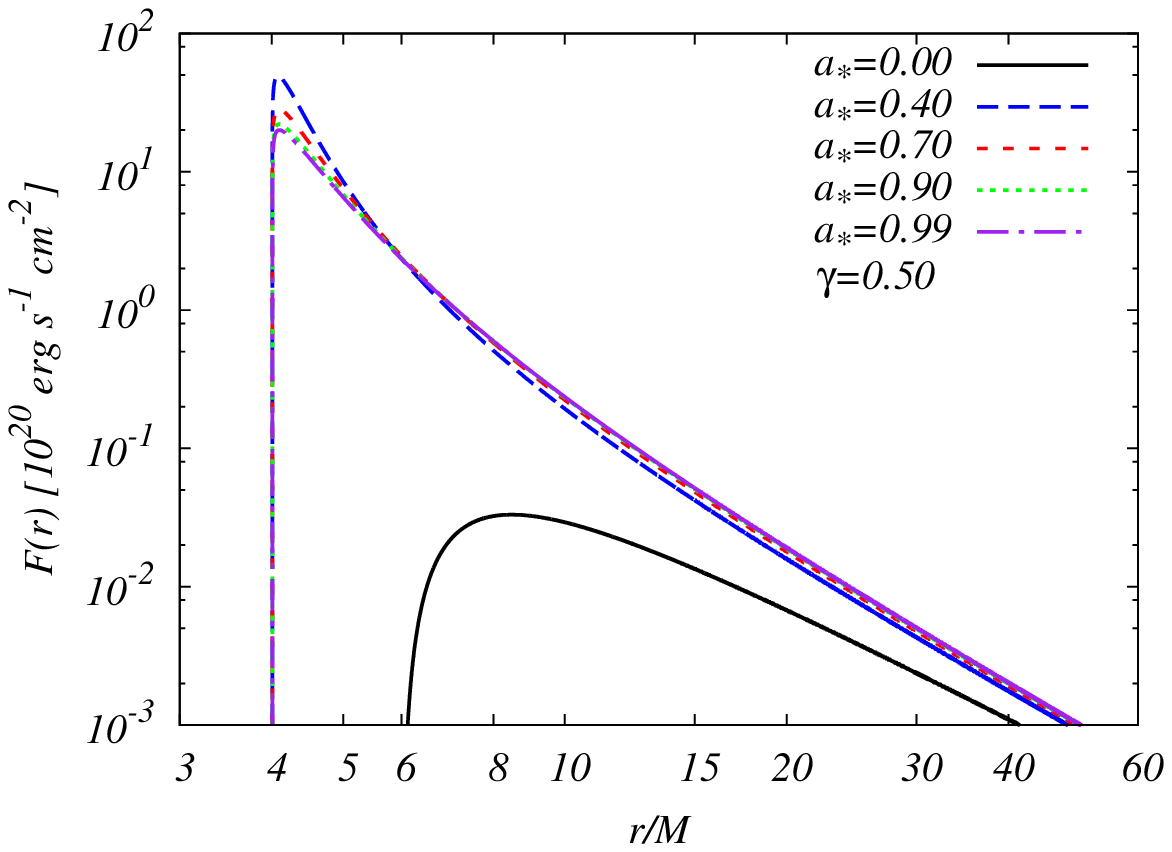}\\
\includegraphics[width=.48\textwidth]{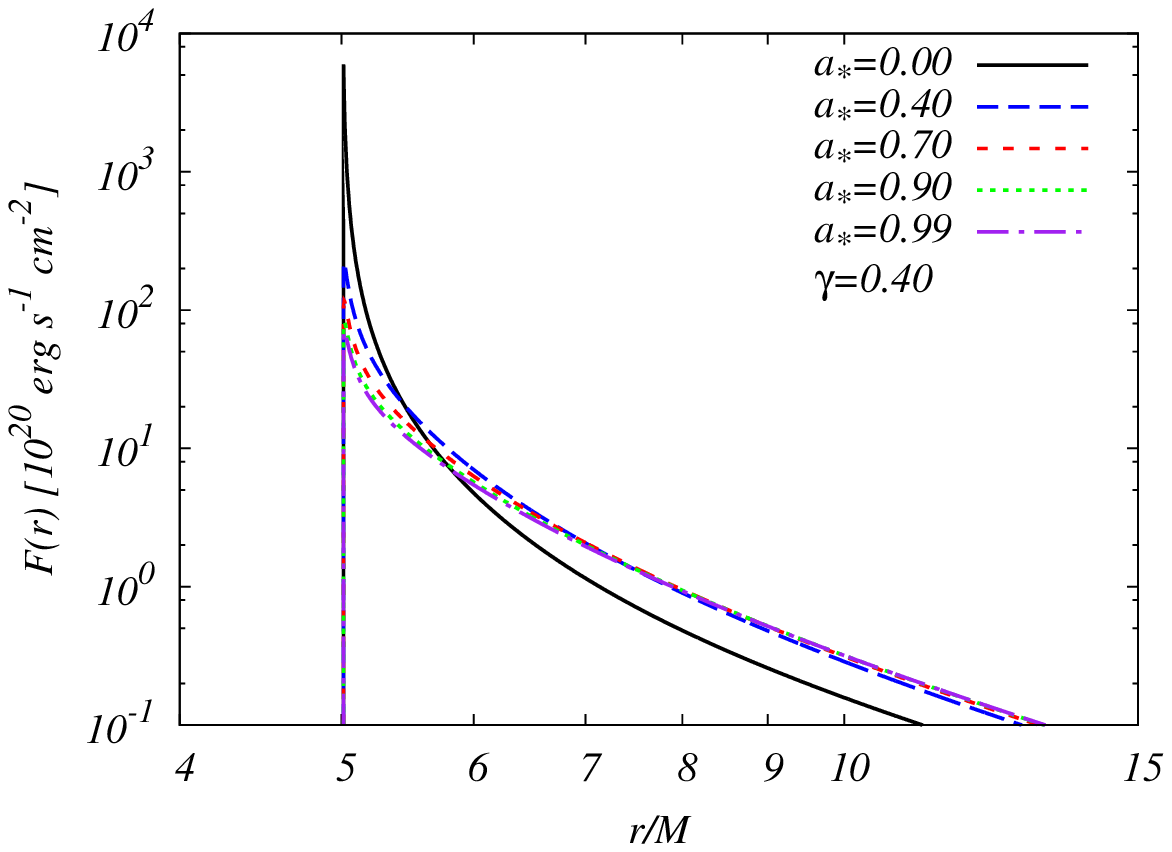}
\includegraphics[width=.48\textwidth]{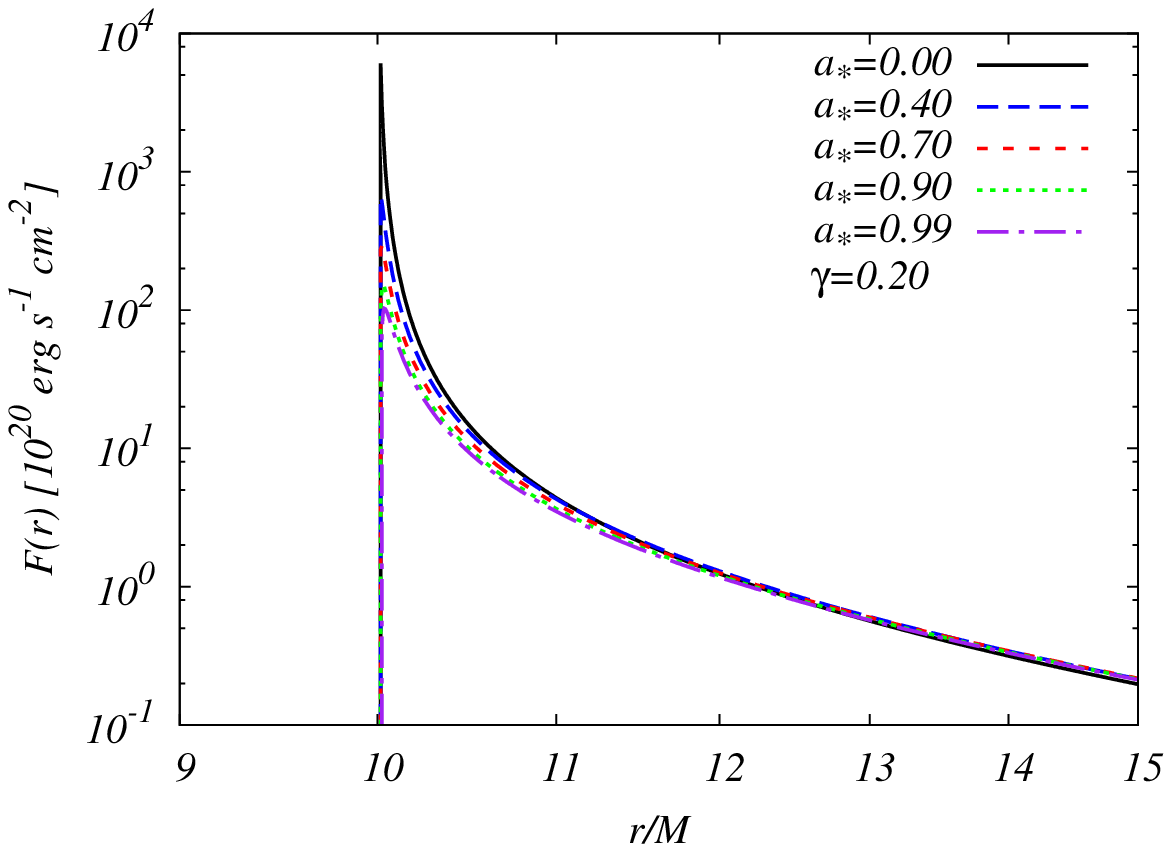}\\
\caption{The energy flux $F(r)$ radiated by the disk around a rotating black hole ($\gamma=1$), and a naked singularity ($\gamma=0.99$, 0.7, 0.5, 0.4 and 0.2), with the total mass $5M_{\odot}$, for different values of the spin parameter $a_*$. The mass accretion rate is set to $10^{-12}M_{\odot}$/yr.}
\label{fig6}
\end{figure}

\end{widetext}

The rest of the panels in Fig.~ \ref{fig6} shows the flux distribution of the thermal radiation of the accretion disks around naked singularities. Depending on the values of the scalar charge parameter $\gamma$, and of the spin parameter $a_*$, naked singularities and black holes could exhibit either similar, or rather different properties.
For $\gamma=0.99$, we obtain flux profiles similar to the corresponding cases of the black holes, with the same spin, provided that the equation $V_{,rr}=0$ has a real solution. The top right hand panel in Fig.~\ref{fig2} shows that for $a_*=0$, 0.4 and 0.7 there exists a radius where the second order derivative of the effective potential vanishes, whereas for $a_*=0.9$ and 0.99 $V_{,rr}<0$ in the whole spacetime. Hence for $\gamma=0.99$, $F(r)$ has similar characteristics to the flux profiles derived in the case of the black holes for $a_*=0$, 0.4 and 0.7.,  but the flux distribution around the black hole and the naked singularity is rather different for higher spin values.

 In the slowly rotating case, when the inner edge of the disk is located at the marginally stable orbit, the radial profiles of $F(r)$ have the same shape in both the top left and the right hand side plots, and the flux maxima are somewhat smaller for the black holes as compared to flux maxima for the naked singularity.
This difference is enhanced by the fast rotation of the central naked singularity. Since stable circular orbits around the naked singularity with $a_*=0.9$ and $0.99$ exist in the whole equatorial plane, the inner edge of the accretion disk reaches the singularity. The radius $r_s=2M/\gamma$ is bigger than $r_{ms}$ for the rotating black holes with the spin of 0.9 or 0.99, and therefore the surface of the accretion disk around the rotating naked singularity is smaller than the one for its Kerr black hole counterpart. Even this smaller disk surface is not the whole area that can radiate thermal photons in the thin accretion disk model, since the left end of the flux profile is pushed to a radius somewhat higher than $r_s$. This is due to the fact that $\Omega_{,r}$ changes its sign at $r_{max}(>r_s)$, as seen in Fig.~\ref{fig4}, and this also changes the sign of the cofactor of the integral in the flux formulae, given by Eq.~ (\ref{F}).  As a result, we obtain negative flux values for the radii between $r_s$ and $r_{max}$, (since the sign of the integral remains the same at $r_{max}$).

The negative flux values involve non-physical states in the framework of the stationary thin disk model. It indicates that the thermodynamical equilibrium cannot be maintained in this region and other forms of the energy and momentum transport become dominant over the radiative cooling such as advection or convection. Then the approximation of the steady-state and geometrically thin disk model breaks down here. Only outside this region, which forms  a very thin annulus between $r_s$ and $r_{max}$ though, one can apply the standard accretion disk scheme. We assume the thin disk exists only in the region for $r<r_{max}$, where the radiation gives the main contribution for energy and momentum transport, and other forms of the transport processes becomes negligible.

Here we consider the radiation properties only for the geometrically thin, relatively cold disk, which is truncated at $r_{max}$, and we discuss the contribution of the hot matter in the region $r<r_{max}$ to the total radiation of the disk-naked singularity system in the next Section.

Thus we assume that the accretion disk emits thermal photons only at radii higher than $r_{max}$, and that for $r>r_{max}$ the photon flux can be computed in the thin disk approximation. Since $r_{max}$ does depend only on $\gamma$, but not of $a_*$, the left edge of $F(r)$ is located always at the same radii, irrespectively of the rotational speed of the naked singularity. In spite of this  reduction in the effective radiant area of the disk surface, as compared with the Kerr black hole case, with the same spin parameters (0.9 and 0.99), the flux maxima is much higher for the rotating naked singularity as compared to the black hole case. The two plots on the top of Fig.~\ref{fig6} show that the rotating naked singularity with $a_*=0.99$ has a disk ten times more luminous than the disk of a Kerr black holes with the same spin. Comparing the rapidly rotating naked singularity with its static case we also see a rise of 3 orders of magnitude in the flux maximum, whereas this rise is only 2 orders of magnitude for the black hole. Thus,  $F(r)$ is much more  sensitive for the variation in $a_*$ in the case of the naked singularity. The maximal flux is somewhat higher for $a_*=0.9$ than for $a_*=0.99$, which indicates that the maximum of $F(r)$ for fast rotation is inversely proportional to the spin parameter.

The fact that the flux maximum is higher for the rotating naked singularity than for the black hole, even if it is integrated over a smaller surface area, is the consequence of the considerable difference in the metric determinant, characterizing the four-volume element in which the radiant flux is measured in the vicinity of the equatorial plane. For Kerr black holes $\sqrt{-g}=r^2$ holds in the equatorial approximation, but from the metric (\ref{gtt})-(\ref{gzz}) of the rotating naked singularity we obtain $\sqrt{-g}=f^{1-\gamma}r^2$. Since the shape factor $f$ vanishes as $r\rightarrow r_s$, the function $f^{1-\gamma}$ has a small value at $r_{max}$, which is close to $r_s$. Then the four-volume element is much smaller for the naked singularity than for the Kerr black hole, and it produces much higher values in the flux integral (\ref{F}), even if the disk properties determining $\Omega$, $\tilde{E}$ and $\tilde{L}$ are similar in the two cases,  as far as we integrate from $r_{max}$ in the case of the naked singularity (e.g., in the two plots in the top in Fig.~\ref{fig4}, the values of $\Omega$ have only a moderate difference for the naked singularity and the black hole).

We obtain similar trends in the characteristics of the radiated flux if we further decrease the values of $\gamma$. By considering the plots for $\gamma=0.7$ or 0.5 in Fig.~\ref{fig2}, one can see that the equation $V_{,rr}=0$ has a real solution only for $a_*=0$ and $a_*=0.4$ ($\gamma=0.7$), or only for the static case ($\gamma=0.5$). For higher values of $a_*$, the second order derivative of the effective potential remains negative in the whole equatorial plane, and the inner edge of the disk always jumps to $r_s$.
This effect is shown by the two middle plots in Fig.~\ref{fig6}, where the flux profiles for the naked singularity can be separated into the two groups (similarly to the case of $\gamma=0.99$), formed by the curves similar to those obtained for black holes ($V_{,rr}=0$ has a real solution), and the curves with higher maxima, located almost at their left edge ($V_{,rr}<0$ holds everywhere).  The high flux values obtained for the second group are again the consequence of the rapidly shrinking volume element in the vicinity of the singularity. We also see that the flux maxima for the latter group is inversely proportional to the spin parameter, but the curves fall more rapidly for lower spin.

\begin{widetext}

\begin{figure}
\centering
\includegraphics[width=.48\textwidth]{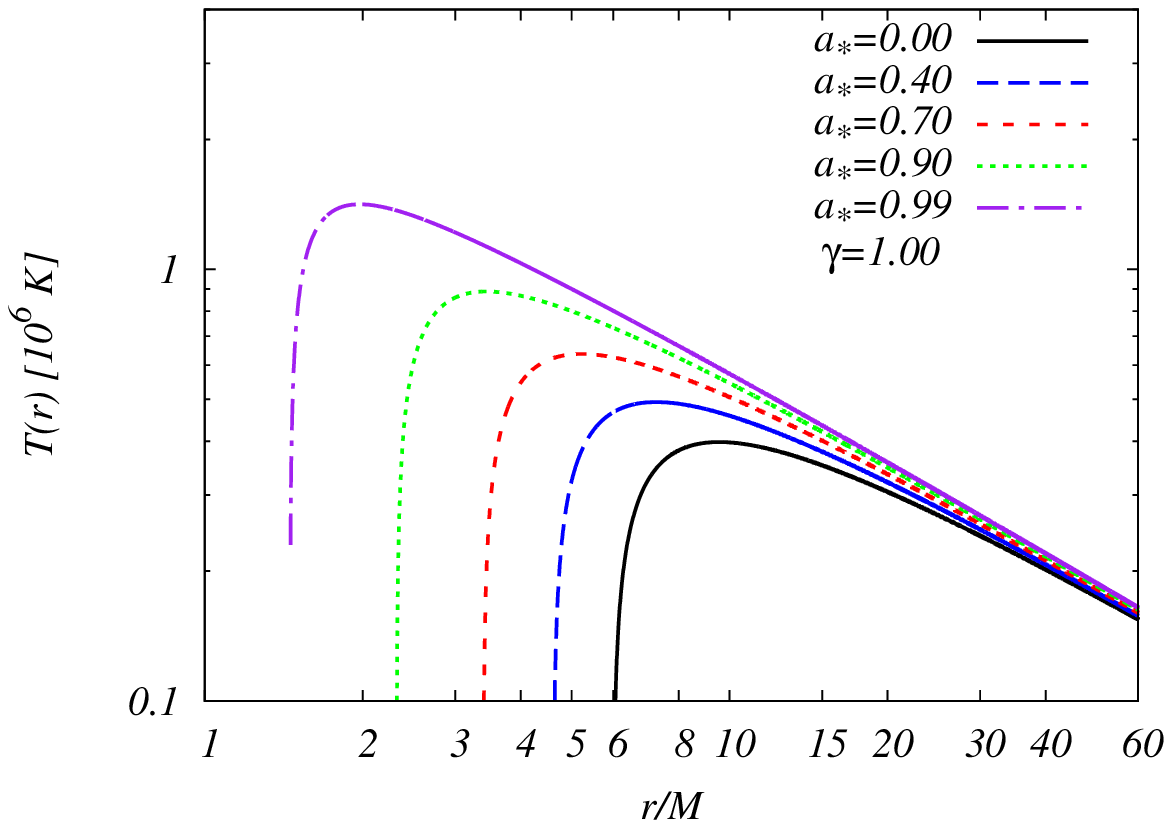}
\includegraphics[width=.48\textwidth]{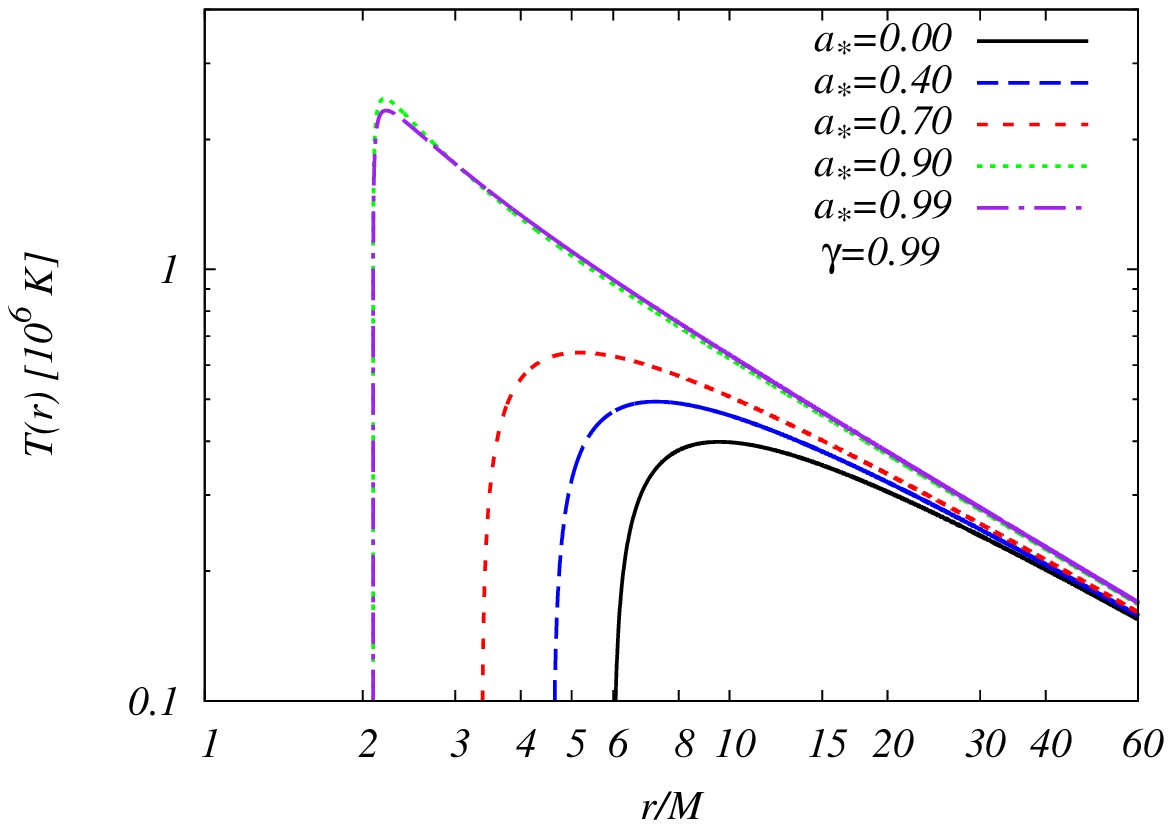}\\
\includegraphics[width=.48\textwidth]{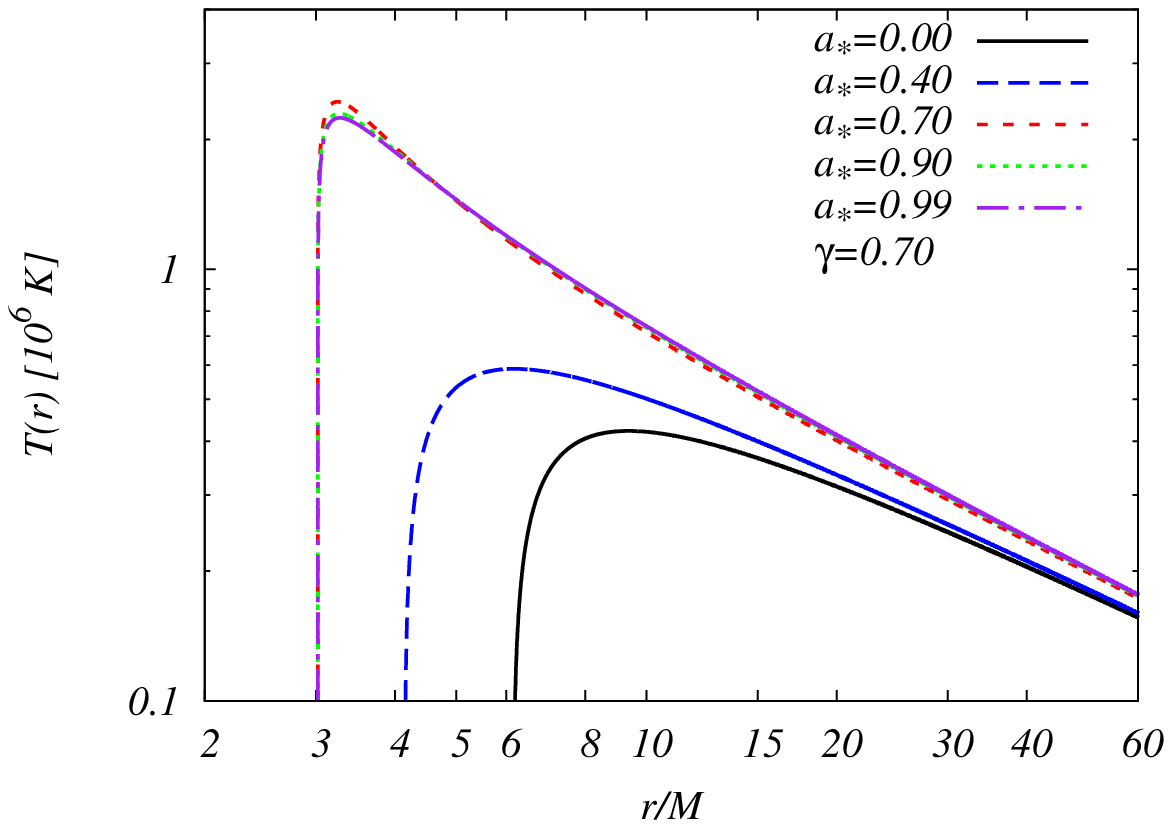}
\includegraphics[width=.48\textwidth]{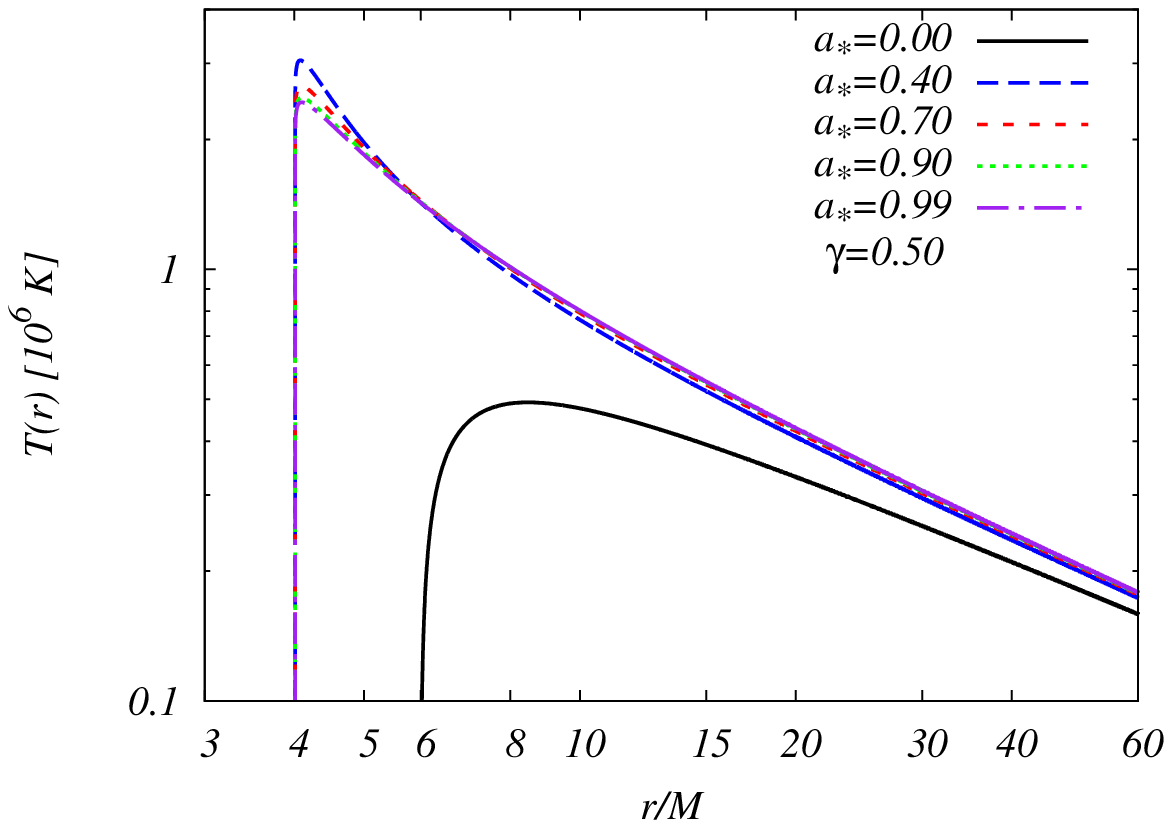}\\
\includegraphics[width=.48\textwidth]{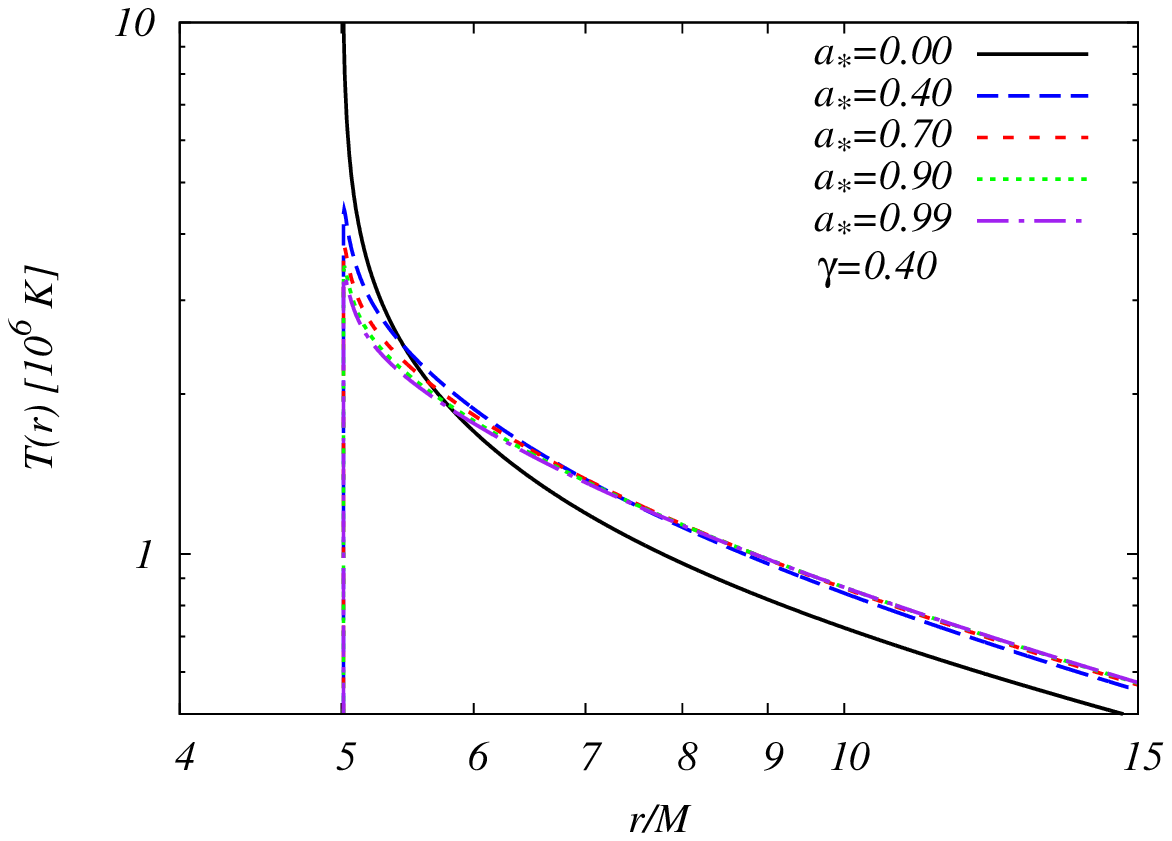}
\includegraphics[width=.48\textwidth]{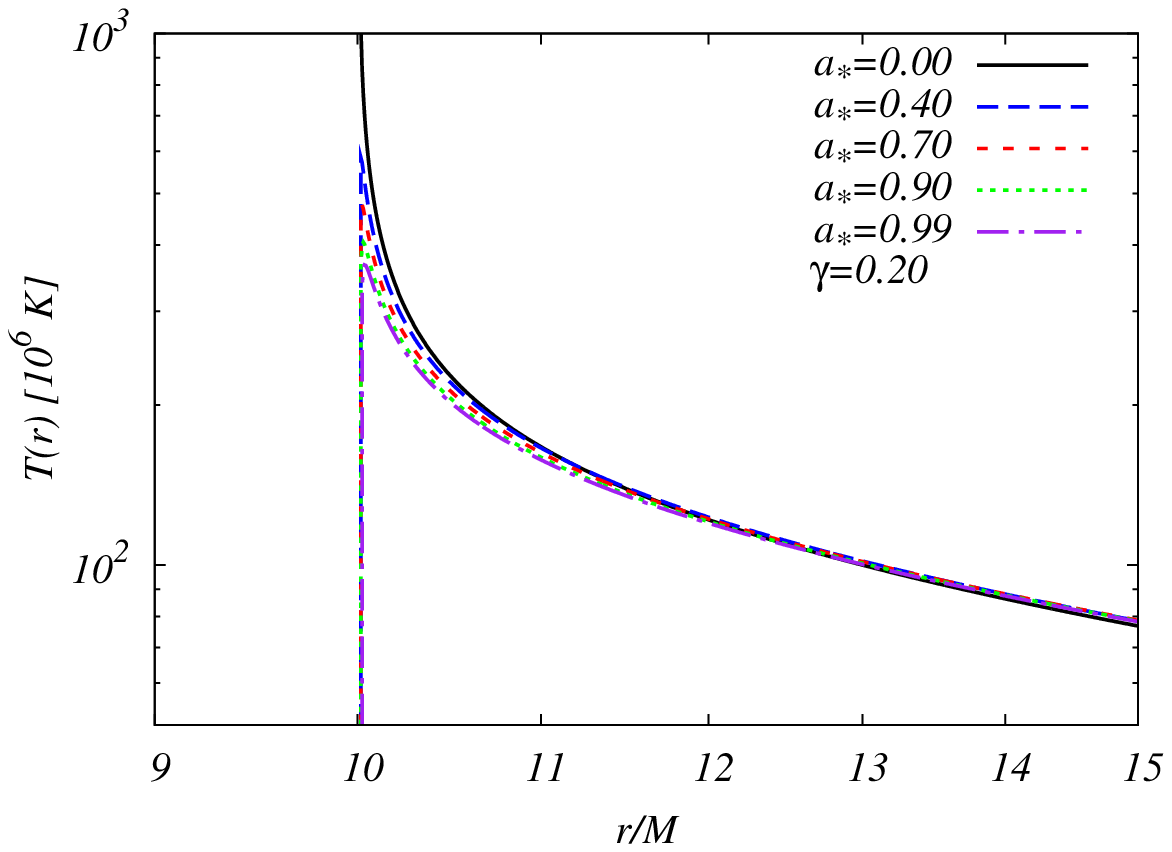}\\
\caption{The temperature distribution on a disk around a rotating black hole ($\gamma=1$), and a naked singularity ($\gamma=0.99$, 0.7, 0.5, 0.4 and 0.2), with the total mass of $5M_{\odot}$, for different values of the spin parameter $a_*$. The mass accretion rate is set to $10^{-12}M_{\odot}$/yr.}
\label{fig7}
\end{figure}

\end{widetext}

Since $\Omega$ approaches zero at the singularity for $1/2<\gamma<1$, the inner boundary of the radiant disk area is located at $r_{max}$, where we obtain finite flux values for $a_*\rightarrow0$. For $\gamma=1/2$, $\Omega$ has a non zero value at $r_s$, and it is a monotonous function in the entire equatorial plane. Therefore, we  cannot use $r_{max}$ as a truncation radius for the thin disk, and the inner edge of the disk would reach the singularity, where the volume element shrinks to zero, and the cofactor of $(-g)^{-1/2}$ in the flux formula (\ref{F}) goes to infinity. Their multiplication for $\gamma=1/2$ causes  $F(R)$ to remain finite at the singularity, as seen in Fig.~ \ref{fig6}. For all non-static cases, plotted in the middle right hand plot in Fig.~\ref{fig6}, the peaks at the left edge of the flux profiles have infinite amplitudes, which cannot represent physical states. The thin diks model is not a good approximation at the inner edge of the disk, as the steady state of the disk cannot be maintained close to the singularity. We can still state that the accretion disk must be extremely bright in its innermost area, but there must be some other physical mechanisms besides the radiative cooling, such as advection, which have an important role in the energy and angular momentum transport.

For $\gamma<1/2$, $\Omega\sim a^{-2}_*$ holds at $r_s$, and $\Omega$  will diverge as $a_*$ approaches zero. Its derivative with respect to $r$  also tends to infinity, also giving, together with the vanishing volume element at $r_s$, an infinite flux value in Eq.~(\ref{F}). This effect is demonstrated by the bottom plots in Fig.~\ref{fig6}. Since $\Omega$ and its derivative is inversely proportional to $a_*$, the rate of the rise of the flux profiles as approaching $r_s$ is also inversely proportional to $a_*$. Thus, the slowly rotating naked singularities have brighter accretion disks than their fast rotating counterparts do have, but in both cases $F(r)$ tends to infinity at the inner edge of the disk.
Without any upper boundary in the flux maxima at the singularity, the physical state of the matter in motion close to the singularity will recede from the thermodynamical equilibrium. The standard accretion disk model cannot be applied at such small distances from the singularity. At higher radii there must exists a region,  where the disk matter can attain thermodynamical equilibrium. If this radius is still close enough to $r_s$, the flux profiles, shown in the bottom plots of Fig.~\ref{fig6}, still resemble the real physical situation, with some uncertainties in the position of their left edges. Obviously, in a physical situation, the amplitudes of the peaks must have some finite values,  but these maxima are still much higher for rotating naked singularities as compared with those calculated for disks rotating around Kerr black holes.

In Fig.~\ref{fig7} we present the temperature distribution of the accretion disk for the same configurations which we have used to study the flux profiles. The disk temperature exhibits a similar dependence on the parameters $\gamma$ and $a_*$ as $F(r)$ does. With decreasing $\gamma$, and increasing $a_*$,  we obtain temperature profiles with much higher and sharper maxima than those for the Kerr black hole. For $\gamma\ge1/2$, there still exist configurations with lower spin, which give radial temperature profiles similar to those obtained for the Kerr black holes. Although these temperature profiles become uncertain in the innermost region of the accretion disk, the disk must still be extremely hot in this region, as compared to the typical disk temperatures obtained for Kerr black holes, with the same spin values.

\begin{widetext}

\begin{figure}
\centering
\includegraphics[width=.48\textwidth]{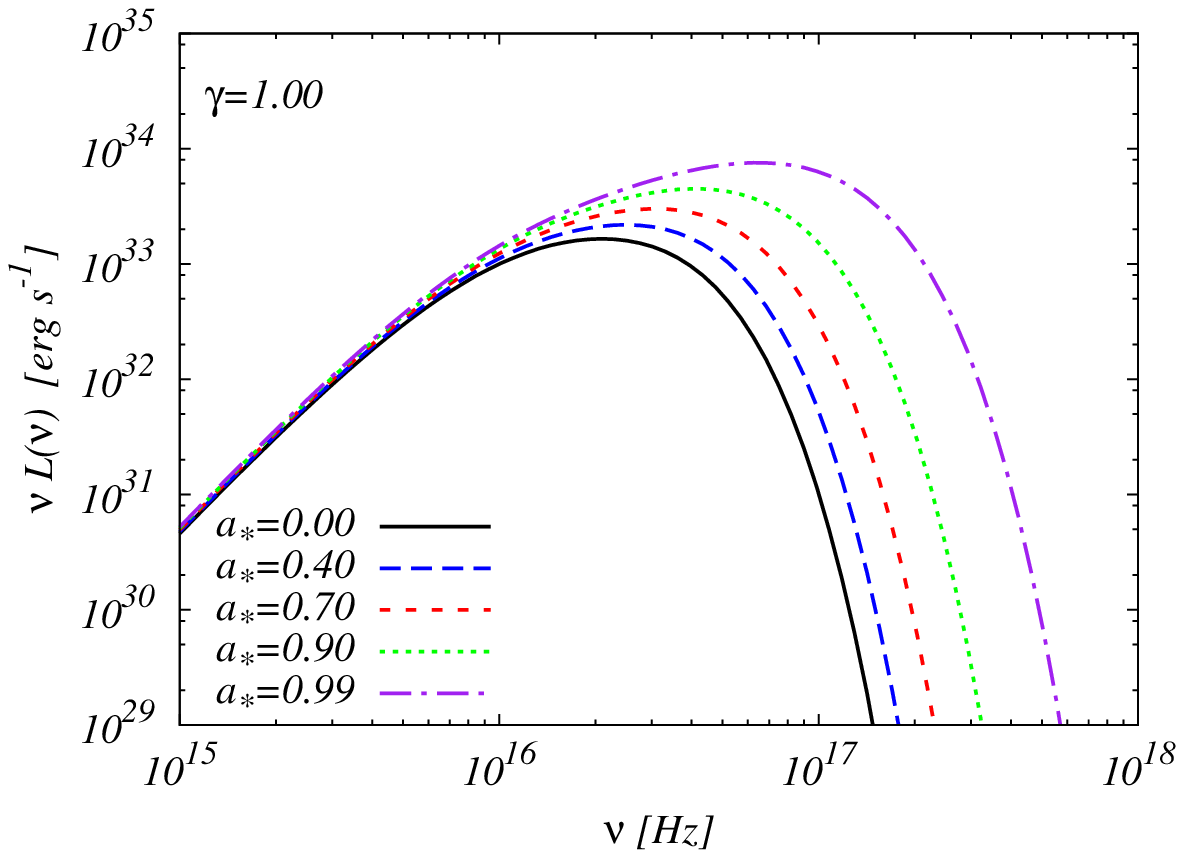}
\includegraphics[width=.48\textwidth]{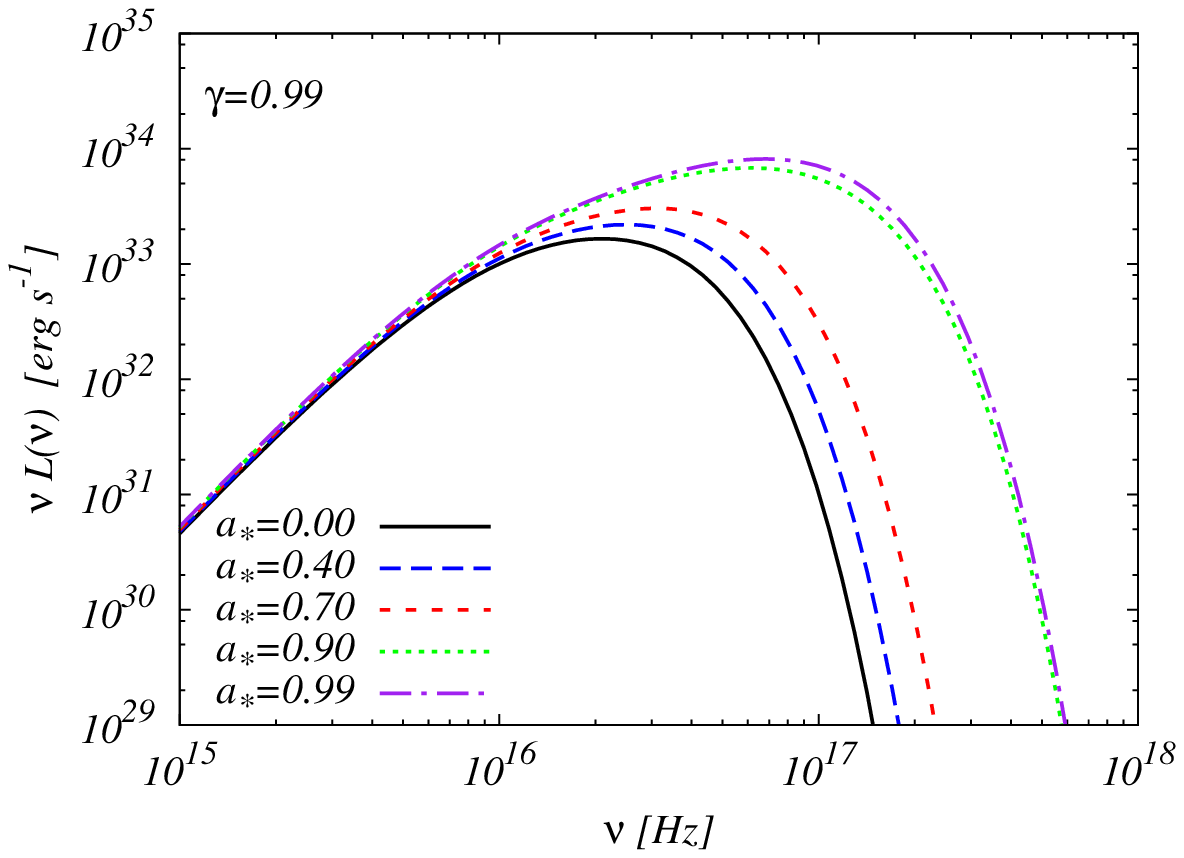}\\
\includegraphics[width=.48\textwidth]{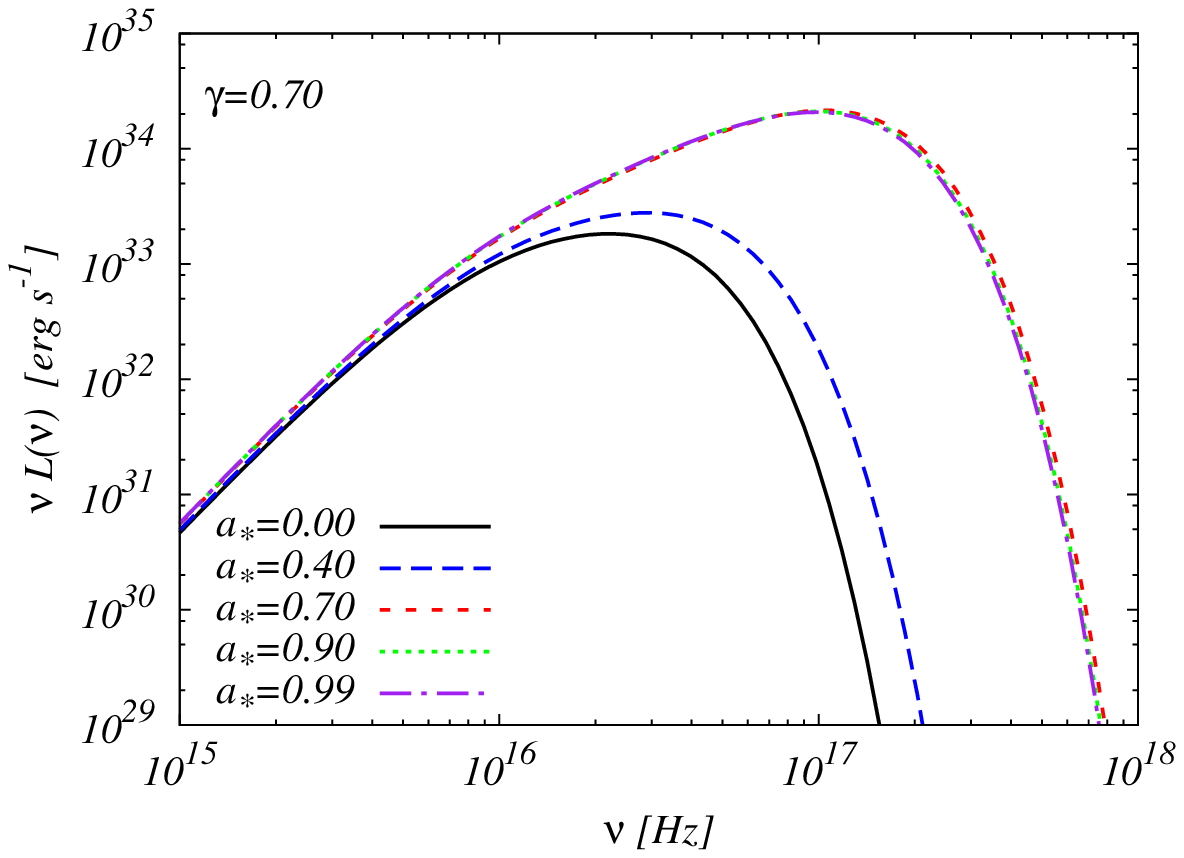}
\includegraphics[width=.48\textwidth]{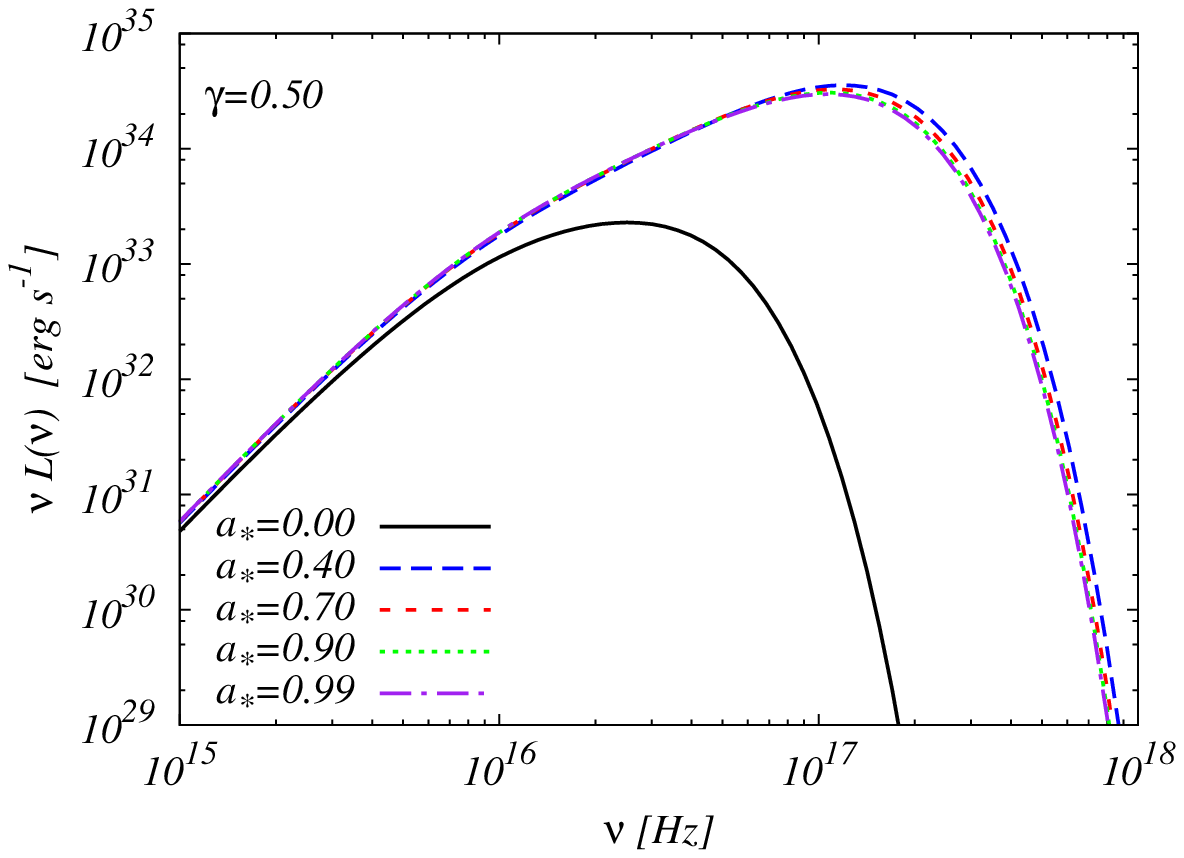}\\
\includegraphics[width=.48\textwidth]{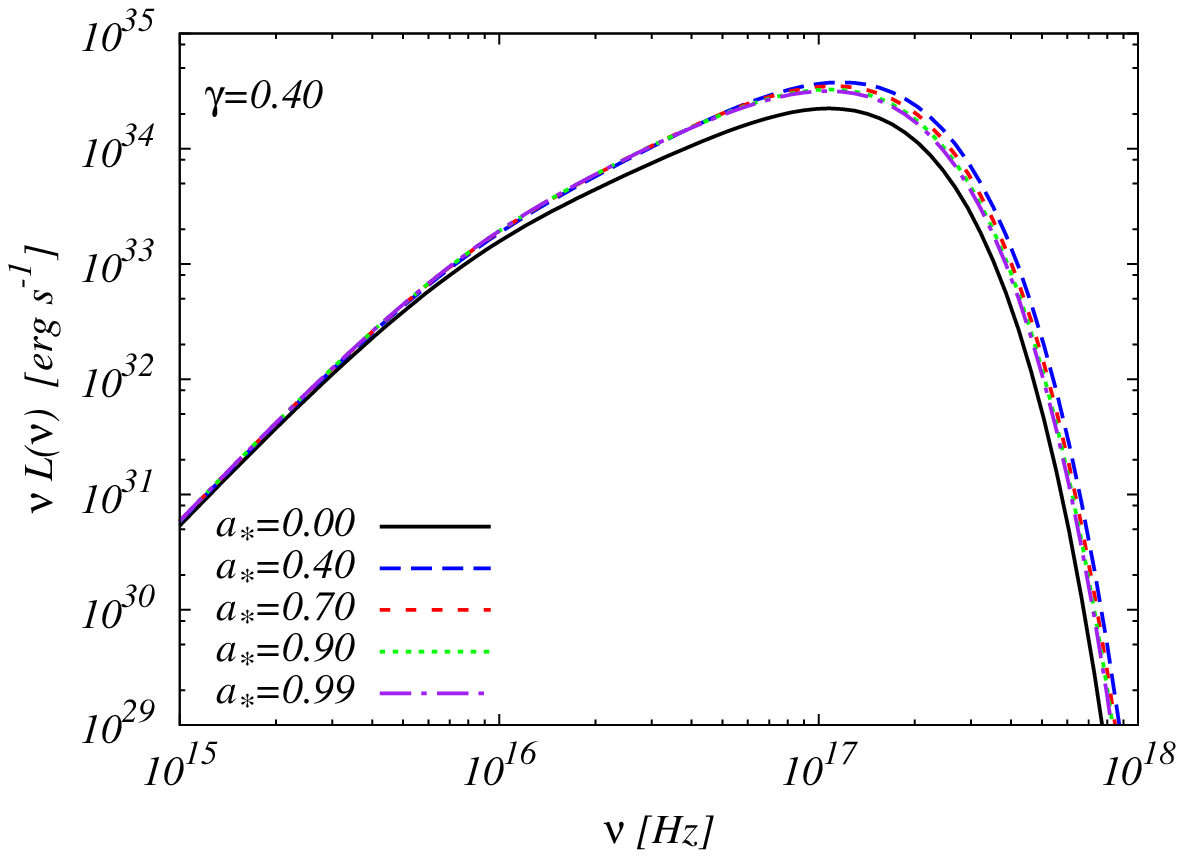}
\includegraphics[width=.48\textwidth]{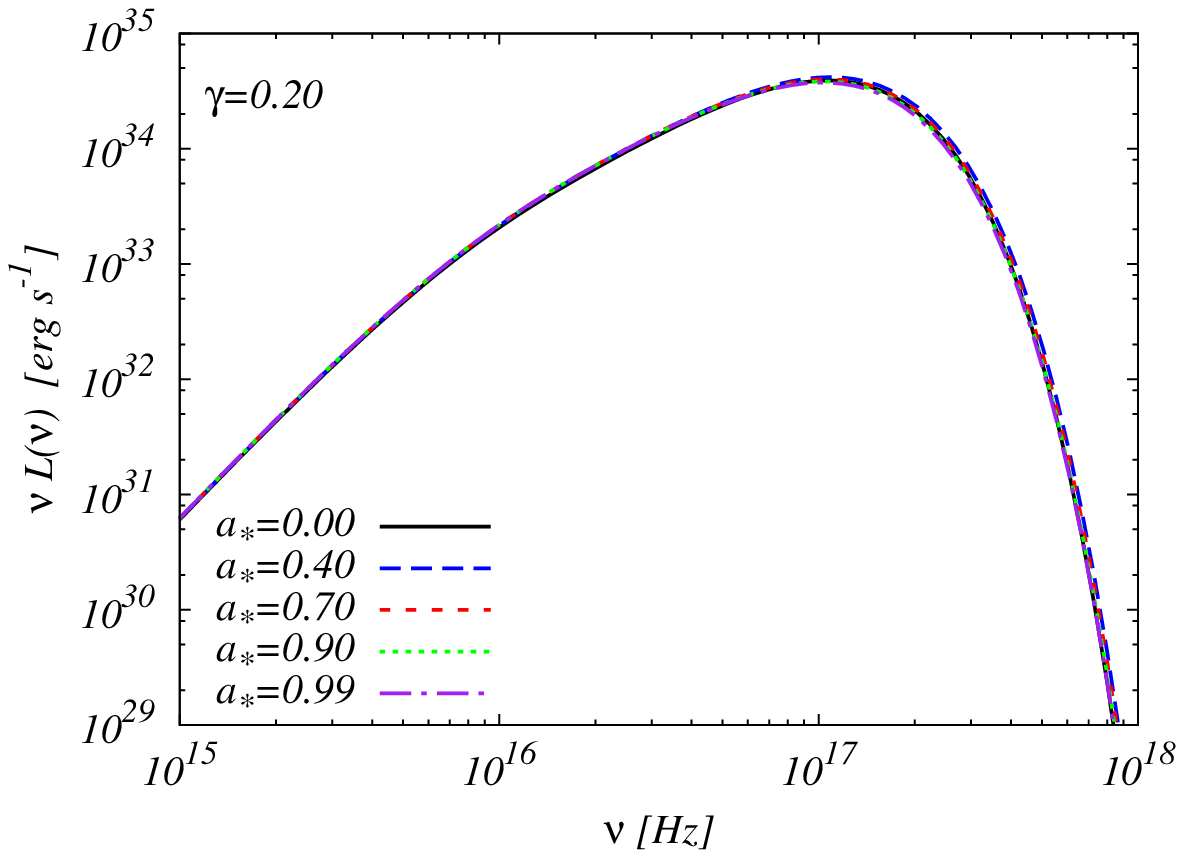}\\
\caption{The accretion disk spectra for a rotating black hole ($\gamma=1$), and a naked singularity ($\gamma=0.99$, 0.7, 0.5, 0.4 and 0.2), with the total mass of $5M_{\odot}$, for different values of the spin parameter $a_*$. The mass accretion rate is set to $10^{-12}M_{\odot}$/yr.}
\label{fig8}
\end{figure}

\end{widetext}

In Fig.~\ref{fig8} we have also plotted, for the same set of values of the parameters $a_*$ and $\gamma$,  the disk spectra, which were calculated from the luminosity formula Eq.~(\ref{L}). These plots show the same trends found in the behavior of the radiated flux distribution $F(r)$, and of the disk temperature in the black hole and naked singularity geometries, respectively. For the Kerr black hole, with $\gamma=1$, the cut-off frequency of the spectra shifts towards higher values, and the maximal amplitudes increase with the increasing spin parameter, i.e, the accretion disks of Kerr black holes become hotter by rotating faster, and they produce a blueshifted surface radiation with higher intensity. In the case of the rotating naked singularity, the two classes already identified in the previous discussion of the radial distribution of $F(r)$ and of $T(r)$ preserve their different natures. We have seen that there is
only a moderate decrease in the disk temperature for the static solution, and for rotating naked singularities with $\gamma=0.99$ and with spin of $a_*=0.4$ and 0.7, as  compared to the Kerr black hole with the same spin values. As a result, the disk spectra exhibit only negligible differences for black holes and naked singularities. For fast rotation ($a_*=0.9$ and 0.99), the accretion disk is much hotter in the area close to the singularity. Hence, the cut-off frequencies shift blueward, and the maximal amplitudes of the spectra are much higher then for the black hole disk spectra. The spectral features for the second group are not sensitive to the variations in the spin,
and the relative shifts in the cut-off frequencies and the spectral maxima are very small, as shown in the bottom right hand plot in Fig.~(\ref{fig8}).

In Table~\ref{tab1} we present the conversion efficiency $\epsilon$ of the accreted mass into radiation for both rotating naked singularities and black holes, for different values of the parameters $a_*$ and $\gamma$. This Table demonstrates that with increasing spin parameter, and decreasing $\gamma$ (increasing scalar charge), the efficiency is increasing to a maximal value, and then it starts to decrease again. The rate of the increase and of the decrease depends on the location of the inner edge of the disk, as can be seen in Eq.~(\ref{epsilon}). The behavior of the two groups, with the inner disk edge at $r_{ms}$ and $r_s$, respectively, is the same in this respect, but $\epsilon$ varies in different ranges for the two groups. For the configurations with inner disk edge located at $r_{ms}$, the increase in $a_*$ causes only a moderate variation in $\epsilon$.  While the efficiency for the slowly rotating case is somewhat greater for the naked singularity, the rapidly rotating black holes have a considerable higher efficiency than the naked singularities, with the same spin parameter, do have.
Comparing the first two lines of the table we see that the static and the slowly rotating configurations have the same efficiencies, whereas the accreted mass-to-radiation conversion mechanism is about four times more efficient for extremely rapidly rotating  black holes then for the fast spinning naked singularities, with $\gamma=0.99$. We find relatively small values in the first two columns of the third line, and in the first column of the fourth line as well.
All the values in the last four lines belong to the second group of naked singularities, for which the inner disk edge is located at the singularity (indicated with the same values in the parenthesis in each line as $r_s$, and which  does not depend on $a_*$). In these cases the accretion disk radiates a great amount of thermal energy, and the values of the efficiency are considerably higher than those found for the first group.
For the static cases the efficiency can reach 50\% for $\gamma=0.4$, but it drops to 40\%, as we decrease $\gamma$ to 0.2. With increasing spin, $\epsilon$ has a very mild increase, reaching the values 70\% ($\gamma=0.5$) and 61\% ($\gamma=0.4$) at $a_*\sim0.4$, and then it starts to decrease for faster rotating singularities. For the naked singularity with the spin of $0.99$ and $\gamma=0.2$,  the efficiency falls to 34\%,  which is still higher than the efficiency for extremely fast rotating Kerr black holes. We can conclude that
there is a range of the spin parameter $a_*$, and of the scalar charge $\gamma$,
where Kerr black holes represent more effective engines
in the conversion of accreted mass to radiation than the naked singularities do.
Nevertheless, we have also found another range of these parameters, where the conversion efficiency for rotating naked singularities is much higher than for Kerr black holes.
\begin{widetext}

\begin{table}[tbp]
\begin{center}
\begin{tabular}{|c|c|c|c|c|c|}
\hline
$\gamma$ & $a_*=0$ & $0.4$ & $0.7$ & $0.9$ & $0.99$ \\
\hline
1.00 & 5.72\% (6.00) & 7.51\% (4.62) & 10.4\% (3.40) & 15.6\% (2.32) & 26.4\% (1.46) \\
0.99 & 5.72\% (6.00) & 7.51\% (4.62) & 10.4\% (3.37) & 6.03\% (2.02) &  5.92\% (2.02) \\
0.70 & 6.03\% (6.15) & 8.92\% (4.14) & 73.4\% (2.86) & 70.8\% (2.86) & 69.7\% (2.86) \\
0.50 & 6.94\% (6.00) & 70.6\% (4.00) & 65.1\% (4.00) & 61.8\% (4.00) & 60.4\% (4.00)  \\
0.40 & 50.7\% (5.01) & 61.3\% (5.01)& 56.4\%  (5.01) & 53.2\% (5.01) & 51.8\% (5.01) \\
0.20 & 40.8\% (10.0) & 41.6\% (10.0) & 38.0\% (10.0) & 35.3\% (10.0) & 34.3\% (10.0)  \\
\hline
\end{tabular}
\caption{The efficiency, measured in percents, and $x^2=r/M$ (values in parenthesis) at the inner edge of the accretion disk for Kerr black holes, and rotating naked singularities, respectively. The value of $\gamma=1$  corresponds to the Kerr black hole.} \label{tab1}
\end{center}
\end{table}
\end{widetext}

Some composite accretion disk models consider a geometrically thick and optically thin hot corona,  positioned between the
marginally stable orbit and the inner edge of the geometrically thin, and
optically thick, accretion disk, where an inner edge is lying at a few gravitational radii (\cite{ThPr75}.
In other type of composite models, the corona lies above and under the
accretion disk, and the soft photons, arriving from the disk, produce a hard
emission via their inverse comptonization by the thermally hot electrons in
the corona \cite{LiPr77}. In both configurations, the disk is
truncated at several gravitational radii - reducing the soft photon flux of
the disk - and the soft and hard components of the broad band X-ray
spectra of galactic black holes were attributed to the thermal radiation of
the accretion disk and the emission mechanism in the corona, respectively.

Although the metric potential gives stable circular geodesic orbits in the entire equatorial plane of the spacetime, it cannot guarantee that the matter in motion can maintain configurations with thermodynamical equilibrium in the whole region, which is the basic assumption of the geometrically thin steady state accretion disk model. A natural limit of the region where the standard accretion disk model applies seems to be the radius $r_{max}$, where the Keplerian rotational frequency becomes inversely proportional to the radial distance from the naked singularity.
This inversion in the $\Omega$ versus $r$ relation indicates that the structure equations of the Novikov-Thorne disk model have non-physical solutions in the region below $r_{max}$. Since this limiting radius depends only on the parameter $\gamma $ and the geometrical mass $M$ via Eq.~(\ref{x2pm}), the lowest radial limit of the validity of the thin disk approximation is not a free parameter, but it is determined by the charge parameter of the massless scalar field $\varphi$.

\section{Observational implications}\label{impl}

An interesting effect, involving the Eddington luminosity for the case of a boson star, was pointed out in \cite{To02}. The Eddington luminosity, a limiting luminosity that can be obtained from the equality of the gravitational force inwards and of the radiation force outwards, is given by $L_{Edd}=4\pi Mm_p/\sigma _T=1.3\times 10^{38}\left(M/M_{\odot}\right)$ erg/s, where $m_p$ is the proton's mass, and $\sigma _T$ is the Thompson cross section \cite{ste}. Since a boson star- as well as any other transparent object - has a non-constant mass distribution, with $M=M(r)$, and therefore the Eddington luminosity becomes a coordinate dependent quantity, $L_{Edd}(r)\propto M(r)$. A similar effect occurs for the case of the naked singularity solution considered in the present paper. For simplicity in the following we will consider only the case of the static naked-singularity. One can associate to the scalar field described by this energy-momentum tensor a mass distribution $M(r)$ along the equatorial plane of the disk, given by $M(r)=4\pi \int_{r_{s}}^{r}T_{0}^{\varphi 0}r^{2}dr=2\pi
\int_{r_{s}}^{r}g^{rr}\varphi _{,r}\varphi _{,r}r^{2}dr$. By using the explicit form of the scalar field we obtain
\begin{equation}
M(r)=\frac{\pi }{2}\frac{M^{2}\left( 1-\gamma ^{2}\right) }{\gamma ^{2}}%
\int_{r_{s}}^{r}\frac{\Delta }{r^{4}}f^{\gamma -3}dr.
\end{equation}
The corresponding coordinate-dependent Eddington luminosity can be obtained
as
\begin{eqnarray}
L_{Edd}(r)&=&\frac{1.37\times 10^{33}\times  M^{2}\left( 1-\gamma
^{2}\right) }{\gamma ^{2}}\times \nonumber\\
&&\int_{r_{s}}^{r}\frac{\Delta }{r^{4}}%
f^{\gamma -3}dr\;{\rm erg/s}.
\end{eqnarray}

The variation of the Eddington luminosity with respect to the coordinate $r$ is represented, for a naked singularity with $M=5M_{\odot}$, in Fig.~\ref{fig9}.

\begin{figure}
\centering
\includegraphics[width=.48\textwidth]{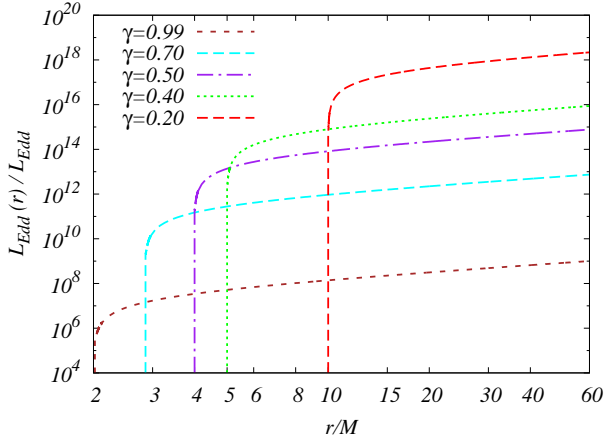}
\caption{Coordinate dependent Eddington luminosity as a function of the radial coordinate $r$ for the case of a static naked singularity with total mass $M=5M_{\odot}$, and for different values of $\gamma $.}
\label{fig9}
\end{figure}

It was argued in \cite{YuNaRe04} that any neutron star, composed by
matter described by a more or less general equation of state,
should experience thermonuclear type I bursts at appropriate mass
accretion rates. The question asked in \cite{YuNaRe04} is whether an
"abnormal" surface may allow such a behavior. The naked singularities may also have such a zero velocity, particle trapping, abnormal surface. The presence of a material surface located at
the singularity implies that energy can be radiated, once matter
collides with that surface.
One can also consider composite accretion disk models as an alternative solution to this problem, and set the truncation radius of the thin disk at $r_{max}$. The hot corona of the composite model can lie in the region between $r_s$ and $r_{max}$, producing hard X-ray spectra via the inverse Compton scattering of photons radiated from the disk and the electron gas in the corona. This hot corona could represent the "surface" of the naked singularity. Thus, at least in principle, naked singularity models,
characterized by high mass, normal matter crusts/surfaces and type
I thermonuclear bursts can be theoretically constructed.

Since naked singularities could be surrounded by a thin shell of matter, the presence of a turning point for matter (the point where the motion of the infalling matter suddenly stops) at the surface of the
naked singularity may have important astrophysical and
observational implications. Since the velocity of the matter at
the naked singularity surface is zero, matter can be captured and deposited on
the surface of the naked singularity.  Moreover, because matter is accreted continuously,
the increase in the size and density of the surface will ignite
some thermonuclear reactions \cite{YuNaRe04}. The ignited reactions are usually
unstable, causing the accreted layer of gas to burn explosively
within a very short period of time. After the nuclear fuel is
consumed, the naked singularity also reverts to its accretion phase,
until the next thermonuclear instability is triggered. Thus, once a thin material surface is formed, naked singularities may undergo a semi-regular series of explosions, called type I thermonuclear bursts, discovered first for X-ray
binaries \cite{Gr76,Tou03}.

The observational signatures indicating the presence of X-ray
bursts from naked singularities are similar to those of
standard neutron stars, and are the gravitational redshift of a
surface atomic line, the touchdown luminosity of a
radius-expansion burst, and the apparent surface area during the
cooling phases of the burst \cite{Ps07}.

If the thermal radiation with wavelength $\lambda _{e}$ emitted by
the thin shell of matter at the surface of the naked singularity has absorption or
emission features characteristic of atomic transitions, these
features will be detected at infinity at a wavelength $\lambda
_{o}$, gravitationally redshifted with a value
$z_{grav}=\left(\lambda _{o}-\lambda _{e}\right)/\lambda _{e}=\left(1-2M/\gamma r\right)^{-\gamma /2}-1$,
where we have assumed, for simplicity, that the radiation emission  takes place in the equatorial plane of the naked singularity. The value of the redshift for a neutron star with mass $M=2M_{\odot}$ and radius $R=10^6$ cm is $z_{NS}=0.56$.  By assuming a naked singularity of mass $M=4\times10^6M_{\odot}$ and radius $R=1.4\times10^{12}$ cm, for $\gamma =1$ we obtain a surface redshift of $z_{grav}=1.55$, for $\gamma =0.99$ we obtain $z_{grav}=1.6057$, while for $\gamma =0.85$, $z_{grav}=9.99$.  Therefore the radiation
coming from the surface of a naked singularity may be highly
redshifted (in standard general relativity the redshift obeys the
constrain $z\leq 2$).

Type I X -ray bursts show strong spectroscopic evidence for rapid
expansion of the radius of the X-ray photosphere. The luminosities
of these bursts reach the Eddington critical luminosity at which
the outward radiation force balances gravity, causing the
expansion layers of the star. The touchdown luminosity of
radius-expansion bursts from a given source remain constant
between bursts to within a few percent, giving empirical
verification to the theoretical expectation that the emerging
luminosity is approximately equal to the Eddington critical
luminosity. The Eddington luminosity at infinity of a naked singularity with a thin surface is given by
\cite{Ps07}
\begin{equation}
L_{Edd}^{\infty }=\frac{4\pi m_{p}r_{s}^{2}}{\sigma _{T}}\left.\left( \frac{\sqrt{%
-g_{tt}}}{\sqrt{g_{rr}}}\frac{d}{dr}\sqrt{-g_{tt}}\right) \right|_{r=r_{s}}
\end{equation}
 For the metric given by Eqs.~(\ref{gtt})-(\ref{gzz}) we obtain
\begin{eqnarray}
L_{Edd}^{\infty } =
\frac{4\pi m_{p}M}{\sigma _{T}}\left[ \frac{\Delta }{r}\left.f
^{3\left(\gamma -1\right)/2 }\right] \right|_{r=r_s}= \nonumber\\
1.3\times 10^{38}\frac{M}{M_{\odot}}\left.\left[ \frac{\Delta }{r}f
^{3\left(\gamma -1\right)/2 }\right] \right|_{r=r_s}.
\end{eqnarray}

For $r\rightarrow 2M/\gamma $, $L_{Edd}^{\infty }\rightarrow \infty$. This shows that the luminosity of naked singularities with thin surfaces can reach much higher values than in the case of standard astrophysical objects.  Finally, we consider the apparent surface area during burst
cooling. Observations of the cooling tails of multiple type I
bursts from a single source have shown that the apparent surface
area of the emitting region, defined as $S^{\infty }=4\pi
D^{2}F_{c,\infty }/\sigma _{SB}T_{c,\infty }^{4} $, where
$F_{c,\infty }$ is the measured flux of the source during the
cooling tail of the burst, $T_{c,\infty }$ is the measured color
temperature of the burst spectrum, $D$ is the distance to the
source and $\sigma _{SB}$ is the Stefan-Boltzmann constant,
remains approximately constant during each burst, and between
bursts from the same source. The color temperature on the surface
of the compact object $T_{c,h}$ is related to the color
temperature measured at infinity by $T_{c,h}=T_{c,\infty }\sqrt{g^{tt}}
$ \cite{Ps07}. By introducing the color correction factor $f_{c}=T_{c}/T_{eff}$, where $%
T_{eff}$ is the effective temperature at the surface, we obtain
\begin{equation}
S^{\infty }=4\pi \left.\left\{\frac{r^{2}}{f_{c}^{4}}\left[ z(r)+1\right] ^{2}\right\}\right|_{r=r_s}.
\end{equation}

Since the radius of the naked singularity  as well as its redshift may be very large quantities, the apparent area of the
emitting region as measured at infinity may be also very large. Hence
all the astrophysical quantities related to the observable
properties of the X-ray bursts, originating at the surface of the
naked singularity can be calculated, and have
finite values on the surface of the naked singularity and at infinity.

\section{Discussions and final remarks}\label{dis}

In the present paper we have considered the properties of the accretion disks that could form around naked singularities, hypothetical singular general relativistic theoretical objects, characterized by the absence of an event horizon. As for the naked singularity we have considered a rotating solution of the Einstein - massless scalar field equations, which reduces to the Kerr solution when the scalar charge tends to 1. As a first step in our study we have investigated the motion of the particles in the gravitational potential of this solution. Depending on the  values of the mass, scalar charge, and of the spin parameter, respectively, there are two types of disks that could exist around naked singularities. For the first type there are marginally stable orbits, located outside the naked singularity, while for the second type the particles can reach, and be in direct contact, with the singularity. Consequently, the properties of the disk radiation are significantly different for these two types of disks. While the first type shows similarities with the Kerr black hole disk, the thermodynamic/electromagnetic properties of the second type could differ by several orders of magnitude from the Kerr disks. A very puzzling result is represented by the behavior of the angular velocity of the particles at the singularity, which is inversely proportional to the spin parameter of the naked singularity. The frame dragging properties of the naked singularity also show significant differences as compared to the Kerr black hole case.

In our approach we have used the thin disk model, which is an idealized physical model. The thin disk model is derived under several physical and geometrical assumptions \cite{PaTh74}. The basic assumptions for the geometrically thin and optically thick model are as follows:
a) The space-time geometry is given by the metric of the rotating naked singularity, and the disk with a negligible self-gravity resides in its central plane.
b) The disk is geometrically thin.
c) There exist a time interval $\Delta t$, small enough for neglecting any change in the parameters of the naked singularity during $\Delta t$, but large enough for measuring the total inward mass flows at any $r$ in the disk, as compared with the total disk mass between $r$ and $2r$.
d) The stress energy tensor of the disk plasma can be algebraically decomposed with respect to the 4-velocity of the plasma.
e) The averaged motion of the baryons over the azimuthal angle and $\Delta t$ can be described as a circular geodesic motion in the equatorial plane.
f) The heat flow within the disk is negligible, except in the vertical direction.
g) The averaged stress-energy propagating to the disk surface is carried by thermal photons, and the photons are emitted from the disk surface vertically on the average.
h) The role of the photons emitted from the disk surface is neglected in the energy and momentum transport between the different regions of the disk. Once any of these conditions is violated, the thin disk model cannot be applied. On the other hand in all the covariant general relativistic formulations of disk models the physical quantities are obtained upon integrations over the four volume element. In our case, the behavior of the volume element near the singular state gives the dominant contribution to the flux, temperature, and spectrum of the disk, and this contribution is much larger than the possible effect on the physical parameters from some improved disk models.

It is generally expected that most of the astrophysical objects
grow substantially in mass via accretion. Recent observations
suggest that around most of the active galactic nuclei (AGN's) or
black hole candidates there exist gas clouds surrounding the
central far object, and an associated accretion disk, on a variety
of scales from a tenth of a parsec to a few hundred parsecs
\cite{UrPa95}. These clouds are assumed to form a geometrically
and optically thick torus (or warped disk), which absorbs most of
the ultraviolet radiation and the soft x-rays. The gas exists in
either the molecular or the atomic phase.
Evidence for the existence of super massive black holes comes from
the very long baseline interferometry (VLBI) imaging of molecular
${\rm H_2O}$ masers in active galaxies, like  NGC 4258 \cite{Miyo95}, and from the astrometric and radial velocity measurements of the fully unconstrained Keplerian orbits for short period stars around the supermassive black hole at the center of our galaxy \cite{Gh08,Gill09}.
The VLBI imaging, produced by Doppler shift measurements assuming
Keplerian motion of the masering source, has allowed a quite
accurate estimation of the central mass, which has been found to
be a $3.6\times 10^7M_{\odot }$ super massive dark object, within
$0.13$ parsecs.

Hence, important astrophysical information can be
obtained from the observation of the motion of the gas streams in
the gravitational field of compact objects. Therefore the study of the accretion processes by compact objects
is a powerful indicator of their physical nature. However, up to
now, the observational results have confirmed the predictions of
general relativity mainly in a {\it qualitative} way. With the present
observational precision, one cannot distinguish between the
different classes of compact/exotic objects that appear in the
theoretical framework of general relativity \cite{YuNaRe04}.
However, important technological developments may allow to
image black holes and other compact objects directly
\cite{Fa00}. Recent observations at a wavelength of 1.3 mm have set a size of microarcseconds on the intrinsic diameter of SgrA* \cite{Do08}. This is less than the expected apparent size of the event horizon of the presumed black hole, thus suggesting that the bulk of SgrA* emission may not be centered on the black hole, but arises in the surrounding accretion flow.
 A model in which Sgr A* is a compact object with a thermally emitting surface was considered in \cite{BrNa06}. Given the very low quiescent luminosity of Sgr A* in the near-infrared, the existence of a hard surface, even in the limit in which the radius approaches the horizon, places a severe constraint on the steady mass accretion rate onto the source: $\dot{M}\leq 10^{-12} M_{\odot}/{\rm yr}$. This limit is well below the minimum accretion rate needed to power the observed submillimeter luminosity of Sgr A*: $\dot{M}>10^{-10} M_{\odot}/{\rm yr}$. Thus it follows that Sgr A* does not have a surface, i.e., that it must have an event horizon. This argument could be made more restrictive by an order of magnitude with microarcsecond resolution imaging, e.g., with submillimeter very long baseline interferometry. Submilliarcsecond astrometry and imaging of the black hole Sgr A* at the Galactic Centre may become possible in the near future at infrared and submillimetre wavelengths \cite{BrLo06}. The expected images and light curves, including polarization, associated with a compact emission region orbiting the central black hole were computed in \cite{BrLo05}. From spot images and light curves of the observed flux and polarization it is possible to extract the black hole mass and spin. At radio wavelengths, disc opacity produces significant departures from the infrared behavior, but there are still generic signatures of the black hole properties. Detailed comparison of these results with future data can be used to test general relativity, and to improve existing models for the accretion flow in the immediate vicinity of the black hole.

With the improvement of
the imaging observational techniques,  it will also be possible to
provide clear observational evidence for the existence of
naked singularities, and to differentiate them from other types of compact
general relativistic objects.

Indeed, in the present paper we have shown that the thermodynamic and
electromagnetic properties of the disks (energy flux, temperature
distribution and equilibrium radiation spectrum) are different for
these two classes of compact objects, consequently giving clear
observational signatures that could help to identify them observationally. More specifically, comparing the energy
flux emerging from the surface of the thin accretion disk around
black holes and naked singularities of similar masses, we have found that
for some values of the spin parameter and of the scalar charge its maximal value is much higher for naked singularities. In fact all the thermodynamical properties of the disks strongly depend
on the values of the spin parameter and on the
scalar charge parameter. These effects are confirmed from the analysis
of the disk temperatures and disk spectra. In addition to this, we have
also shown that for a given range of the spin parameter and of the scalar charge, the conversion efficiency of the accreting mass into radiation of naked singularities is  generally much larger than the conversion efficiency
for black holes, i.e.,  naked singularities provide a much more efficient
mechanism for converting mass into radiation than black holes. Thus,
these observational signatures may provide the possibility of clearly
distinguishing rotating naked singularities from Kerr-type black holes.

\section*{Acknowledgments}

 The work of T. H. was supported by the General Research Fund grant
number HKU 701808P of the government of the Hong Kong Special
Administrative Region.

\end{document}